# DINO4DSTEM: A self-supervised framework for structural discovery in 4D-STEM

Daniel Khaykelson[a],* Lothar Houben[b], Boris Rybtchinski[a]*

[a]Department of Molecular Chemistry and Materials Science, Weizmann Institute of Science, Rehovot 7610001, Israel

[b]Department of Chemical Research Support, Weizmann Institute of Science, Rehovot 7610001, Israel

*Daniel.kh@weizmann.ac.il, Boris.rybtchinski@weizmann.ac.il

**Abstract**

Nanodiffraction using 4D-STEM has become a key technique for quantitative nanoscale structural mapping in materials research, yet interpreting its high-dimensional datasets in structurally complex materials remains a major bottleneck. Existing analysis workflows typically rely on structural models, manual annotation, predefined classes, or sample-specific heuristics, limiting their ability to characterize heterogeneous complex materials. Here, we introduce DINO4DSTEM, a self-supervised machine learning framework that automatically discovers structurally meaningful states directly from raw diffraction data. Without structural models, manual labels, a predefined number of classes, or system-specific parameter tuning, the framework learns representations that organize diffraction patterns by their intrinsic structural similarities, transforming large collections of low-dose measurements into quantitative nanoscale structure maps. Across diverse datasets, DINO4DSTEM consistently identifies the dominant structural degrees of freedom, providing segmentation without human supervision. We applied the framework to the crystallization of indomethacin, a beam-sensitive, multidomain pharmaceutical system, revealing that crystallinity emerges from a partially ordered precursor and spans a continuous spectrum of structural order. The discovered nanoscale structural states are mapped to reveal the evolution of order across the specimen quantitatively. By replacing task-specific analysis with general self-supervised representation learning, DINO4DSTEM provides a broadly applicable framework for quantitative nanoscale structural mapping in complex materials, enabling the discovery of emergent structural organization in heterogeneous, beam-sensitive systems.

## 1. Introduction

Many materials exhibit evolving nanoscale structural heterogeneity, making its mapping a fundamental challenge. In molecular materials, crystallinity, polymorphism, and defect distributions determine properties ranging from pharmaceutical solubility to charge transport and optoelectronic performance[1,2]. Despite its importance, the emergence of structural order remains incompletely understood because many systems evolve through complex free-energy landscapes that contain partially ordered precursors, transient intermediates, local heterogeneities, and competing polymorphic pathways, rather than following classical nucleation[3,4]. Resolving these heterogeneous structural landscapes requires quantitative nanoscale mapping methods, such as electron microscopy, which presents the additional challenge of handling beam-sensitive materials.

Nanodiffraction using four-dimensional scanning transmission electron microscopy (4D-STEM) has emerged as a powerful low-dose approach for quantitative nanoscale structural mapping in complex materials[5]. By recording a complete diffraction pattern at every probe position with fast direct-electron detectors, energy filters, and cryogenic stages, 4D-STEM enables structural characterization of beam-sensitive materials at electron doses that were previously inaccessible[6]. However, this capability shifts the primary challenge from data acquisition to data interpretation. A single experiment yields thousands to millions of weak diffraction patterns, whose structural information must be extracted and organized into physically meaningful maps of local order. Existing analysis approaches remain tailored to specific experimental conditions or prior assumptions[7]. Peak-finding and orientation-mapping methods perform well when diffraction spots are sharp and structural models are available[7], while matrix factorization and clustering approaches typically require user-defined parameters or assumptions regarding the number of structural classes[8]. Supervised machine learning methods, including Segment Anything Model (SAM)[9], achieve accurate segmentation but rely on manual annotation or sample-specific tuning. Unsupervised learning approaches have also been explored[8,10–12], yet conventional autoencoder-based methods generally fail to account for rotational symmetry, separating diffraction patterns from identical structures viewed at different orientations into distinct clusters[13]. As a result, no general framework currently enables quantitative, automated interpretation of diverse 4D-STEM datasets without prior structural knowledge or expert intervention.

Here we introduce DINO4DSTEM, a self-supervised framework that automatically discovers structurally meaningful states directly from raw 4D-STEM diffraction data, without structural models, manual annotation, predefined classes, or sample-specific parameter tuning. By learning orientation-invariant structural descriptors, while preserving physically meaningful scattering information, the framework automatically groups crystallographically equivalent diffraction patterns and generates quantitative nanoscale maps of structural organization across diverse datasets. We validate the approach on simulated diffraction data, where it is invariant to viewing direction and in-plane rotation, and on experimental organic materials, where it reproduces expert-guided segmentation without human supervision.

We then apply DINO4DSTEM to the crystallization of indomethacin, a pharmaceutically important molecular crystal exhibiting multiple polymorphs whose structural evolution has remained difficult to resolve experimentally[14–17]. The framework reveals that crystallinity emerges from a partially ordered precursor, spans a continuum of structural order, and forms an unexpectedly intricate nanoscale landscape. These results establish DINO4DSTEM as a general instrument for quantitative nanoscale structural mapping from nanodiffraction data, providing new insights into the evolution of structural order in beam-sensitive molecular materials.

## 2. Results and Discussion

### 2.1. A self-supervised classifier adapted to nanodiffraction imaging.

A single 4D-STEM scan produces tens of thousands of diffraction patterns, rendering manual classification into structural groups impractical and subjective. Patterns from the same phase may vary owing to thickness or background scattering, while weak reflections are often interpreted inconsistently. DINO4DSTEM eliminates this manual classification step through self-supervised learning, automatically grouping diffraction patterns based on their intrinsic similarities, without human labeling or predefined structural classes (Figure 1). It builds on DINO[18], a self-supervised method developed for natural images: two slightly altered copies of each diffraction pattern are shown to the network, which is trained to describe both the same way, so that across the whole dataset it learns what makes patterns genuinely alike and related patterns fall into the same class (See Experimental Section and SI for details). The output consists of a map locating each class in the sample, and its average diffraction pattern.

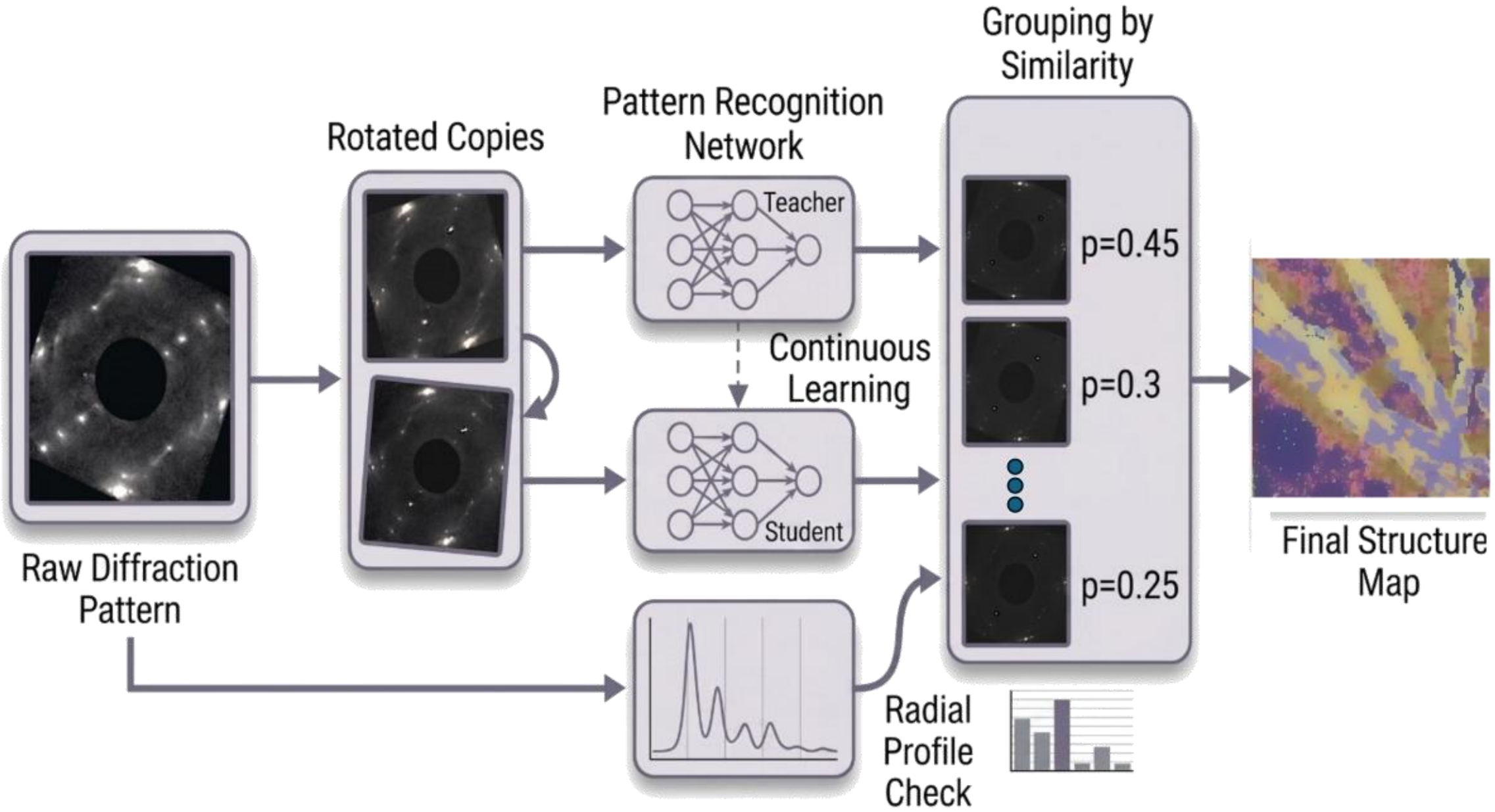

**Figure 1. A label-free self-supervised classifier for 4D-STEM.** A 4D-STEM scan records one nanobeam diffraction pattern at each probe position. Each pattern is shown to the network as two augmented views, which a student

encoder and a slowly updated teacher encoder map into a shared representation; training drives the two views together, and a set of prototypes groups the embeddings into classes. Taking the most likely prototype at each position gives the class (domain) map. The augmentation that sets crystal orientation is the in-plane rotation, so the classes are invariant to orientation. Training uses no labels and no preset number of classes; the objective combines self-distillation with the one-dimensional radial clustering loss. The output panel is a measured DINO4DSTEM class map (indomethacin). Encoder depth, teacher update, and loss weights are given in the Experimental Section.

We adapted the self-supervised framework for diffraction analysis in several key aspects (Figure 1). Because diffraction patterns are relatively simple, we employ a lightweight neural network that enables efficient training and inference. To maintain a direct link to the underlying diffraction physics, a radial scattering term anchors each learned class to its azimuthally averaged scattering profile (Fig. 1). Rotational invariance is achieved by presenting rotated views of the same diffraction pattern during training, enabling crystallographically equivalent patterns at different in-plane orientations to be assigned to the same structural state. Finally, the framework automatically determines the number of occupied structural classes by initializing with a generous upper bound and retaining only those supported by the data, eliminating the need for user-defined cluster numbers. A single set of hyperparameters was established once and applied unchanged across all datasets. Further implementation details and ablation studies are provided in the Experimental Section and Supplementary Figs. S1 and S2.

## 2.2. Benchmarking

We first established the physical significance of the learned structural classes using two benchmark datasets. As a first benchmark, we analyzed simulated polycrystalline $WS_2$ with three known crystal orientations ([100], [110], and [001]), with each domain additionally subjected to arbitrary in-plane rotations, which DINO4DSTEM is designed to be invariant to. The framework recovered the three ground-truth viewing directions with 98.9% accuracy and correctly assigned all in-plane rotations to their corresponding structural classes. In contrast, rotation-variant methods, such as non-negative matrix factorization (NMF) combined with k-means clustering, which are the go-to unsupervised clustering approaches[8], fragmented each viewing direction into multiple spurious classes (Supplementary Figs. S3–S6). We next evaluated DINO4DSTEM on experimental nanodiffraction data from NaPHI, a layered photocatalyst whose flakes exhibit diffuse, line-like scattering that cannot be analyzed by conventional peak-identification methods[19]. In our previous work, these data were segmented into physically meaningful diffraction classes using expert-guided Segment Anything (SAM) analysis[19]. Without human annotation or supervision, DINO4DSTEM reproduced this reference with high fidelity, recovering the line-scattering domain with accuracy comparable to the supervised approach (Fig. 2, spatial intersection-over-union of 0.74, versus 0.77 for polar NMF). The framework also distinguished the NaPHI flakes from the carbon support and resolved finer, physically distinct subdomains that were not identified by other methods (Fig. 2 and Supplementary Figs. S7–S9). Notably, these results were obtained using the same hyperparameters across all datasets, without manual feature selection or dataset-specific tuning, whereas NMF required an additional clustering step whose outcome depended on the chosen algorithm. Together, these complementary benchmarks demonstrate that DINO4DSTEM learns physically meaningful structural classes that generalize across both simulated and experimental 4D-STEM datasets.

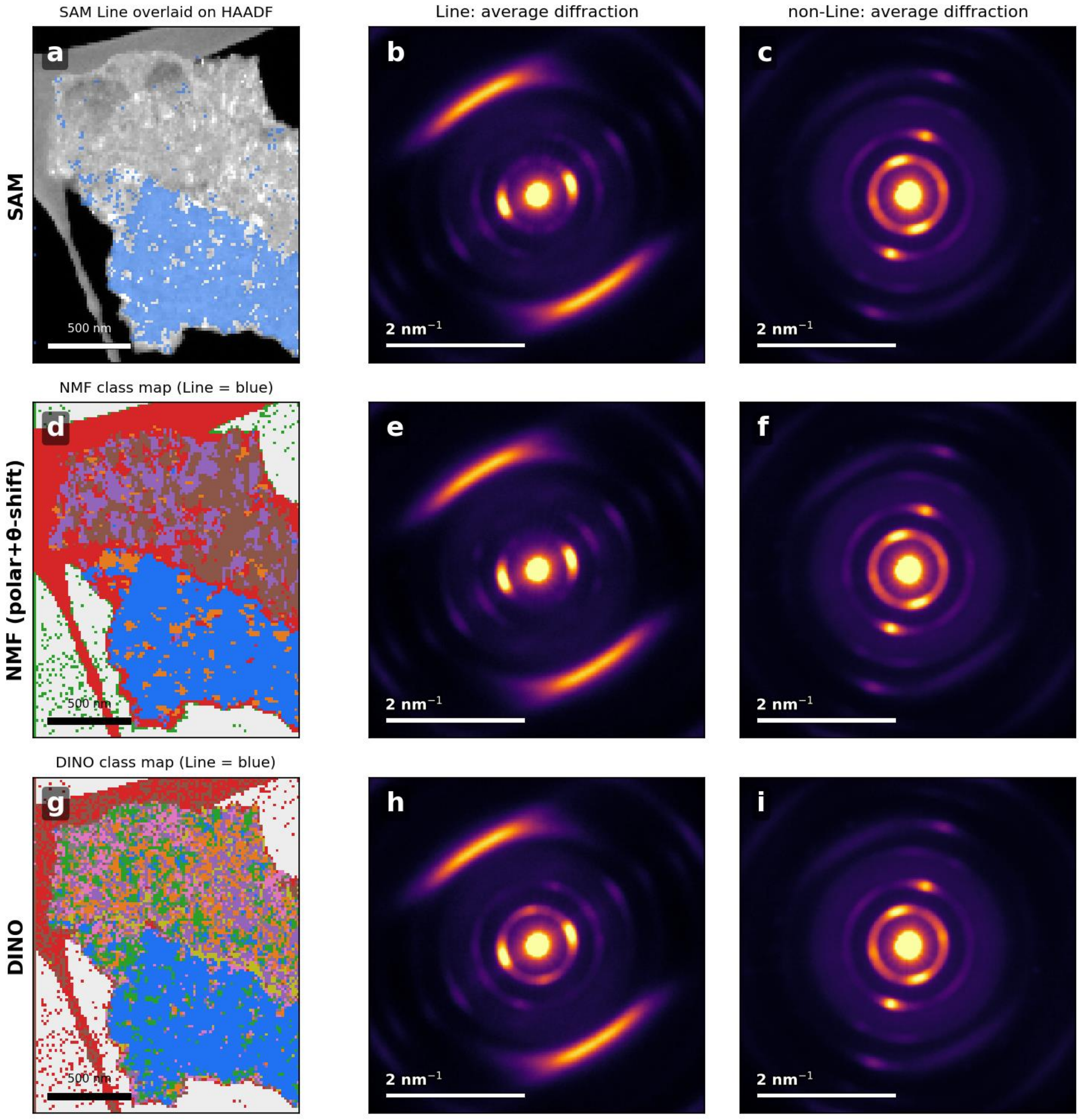

**Figure 2. Validation against an independent expert segmentation (NaPHI).** Rows: SAM (reference), rotation-invariant polar-theta-shift NMF, and DINO4DSTEM. Column 1 shows the oriented line domain: SAM as a blue overlay on the virtual HAADF image (a), and NMF (d) and DINO4DSTEM (g) as full class maps with the line cluster forced to the same blue and the remaining classes in distinct colours. Columns 2 and 3 show the class-averaged diffraction of the line and non-line regions. All three methods recover the line domain (spatial intersection-over-union against SAM of 0.77 for NMF and 0.74 for DINO4DSTEM; because the line is a diffuse azimuthal feature, spatial overlap rather than a radial-profile correlation is the meaningful measure). DINO4DSTEM additionally resolves finer, physically distinct sub-domains, namely thickness variation and a measurement-induced interface, analysed per class in Figure S7, with the full set of NMF classes and their average diffraction in Figure S8 and the line/rest interface domain in Figure S9.

## 2.3. Mapping crystallinity in indomethacin thin films

To demonstrate the capability of DINO4DSTEM for resolving complex structural evolution, we applied the framework to the crystallization of indomethacin. Approximately 150-nm-thick amorphous indomethacin films were prepared by physical vapor deposition[20], followed by annealing at 70 °C under nitrogen to induce crystallization (Fig. 3). DINO4DSTEM was then used to analyze 4D-STEM datasets acquired from partially crystallized films (Fig. 4). Virtual HAADF images reveal crystalline needles emerging from a less ordered matrix (Fig. 4a,f,k).

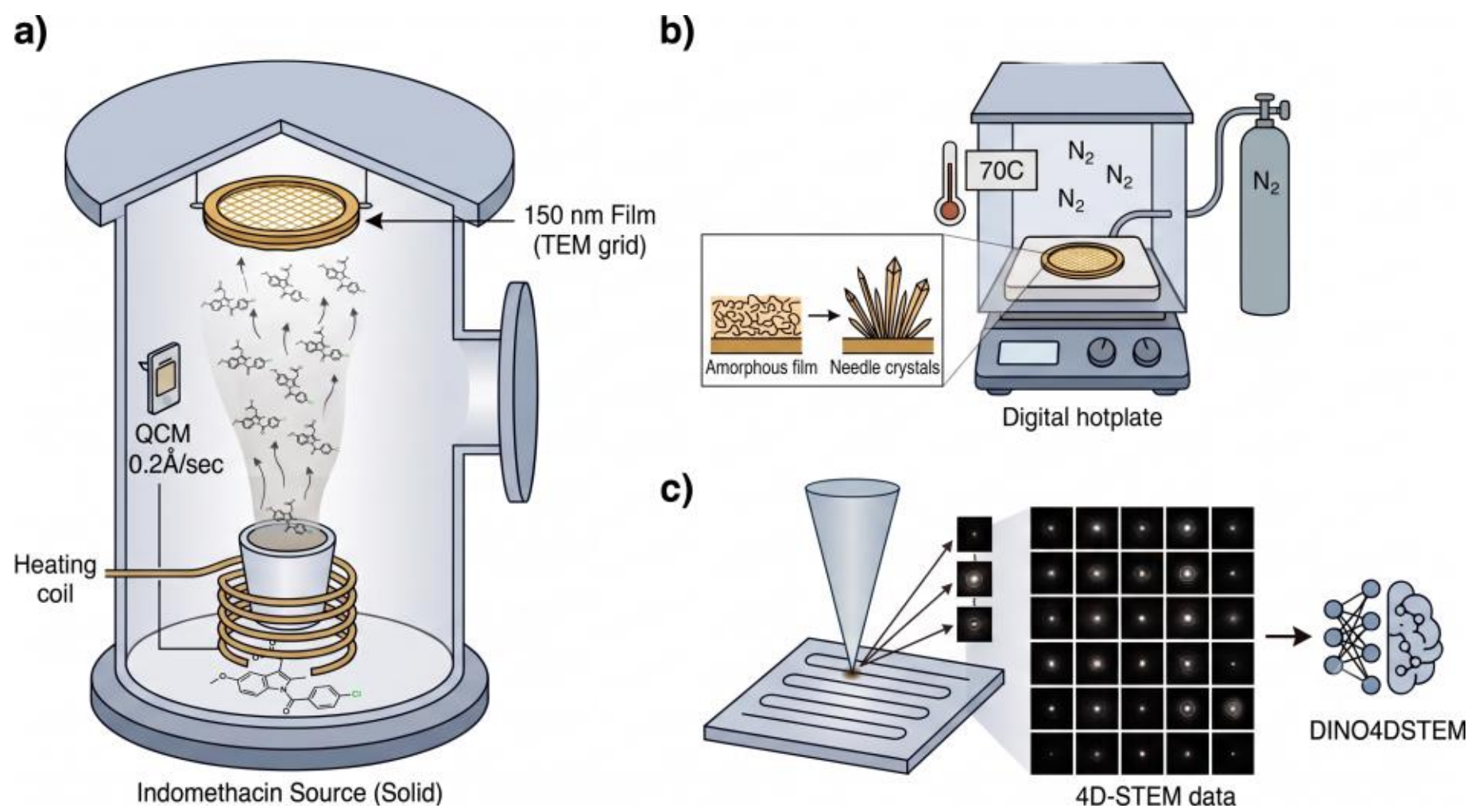

**Figure 3. Overview of the indomethacin crystallization experiment and 4D-STEM acquisition. a)** Physical vapor deposition of indomethacin. The solid drug is sublimed from a resistively heated crucible in vacuum and condenses onto a lacy carbon-coated gold TEM grid suspended above the source, forming an amorphous film approximately 150 nm thick; the deposition rate is monitored by a quartz-crystal microbalance (QCM) and held at 0.2 Å $s^{-1}$. **b)** Ex-situ crystallization. The as-deposited amorphous film is annealed on a digital hotplate at 70 °C under a nitrogen atmosphere, driving the transformation from a featureless amorphous film to crystalline needles (inset); the partially crystallized films that arrest along this pathway are the subject of the present analysis. **c)** Energy-filtered 4D-STEM acquisition. A focused nanobeam is rastered across the film, and a full nanobeam diffraction pattern is recorded at every probe position, producing a four-dimensional dataset of tens of thousands of weak, low-dose patterns fed into DINO4DSTEM for label-free classification.

We compared DINO4DSTEM with polar NMF, which requires an additional clustering step to convert its loadings into a structural map (Fig. 4). Two key differences emerge. First, the NMF result is not uniquely defined because it depends on the clustering algorithm. Applied to the same loadings, k-means, agglomerative, and Gaussian mixture clustering produce different partitions, with adjusted Rand indices (a chance-corrected measure of partition agreement, where 0 indicates random agreement and 1 identical partitions) ranging from 0.15 to 0.69[21] (Supplementary Fig. S10). Consequently, the inferred domain boundaries depend not only on the diffraction data but also on the clustering strategy. In contrast, DINO4DSTEM directly produced a unique structural map without requiring a separate clustering step or algorithmic choice. Its agreement with any individual NMF partition is therefore necessarily modest, reflecting comparison with one member of a family of equally plausible partitions rather than a single well-defined reference.

The result is a unique domain map of a sample whose structure is otherwise hard to access (Figure 4). DINO4DSTEM reconstructs the field of view consistently across the three indomethacin regions that differ in order, most clearly where the crystalline needles give the strongest signal. DINO4DSTEM needs no per-sample tuning and no visible morphological reference, hence the method reduces an ambiguous, operator-dependent clustering task to a single automated step, which is the basis for the physical analysis that follows. This is the payoff of the approach: from one fixed set of settings DINO4DSTEM learns the relevant structural axis directly from the data and returns a single, deterministic and physically meaningful map, complementing peak-finding and NMF-based clustering.

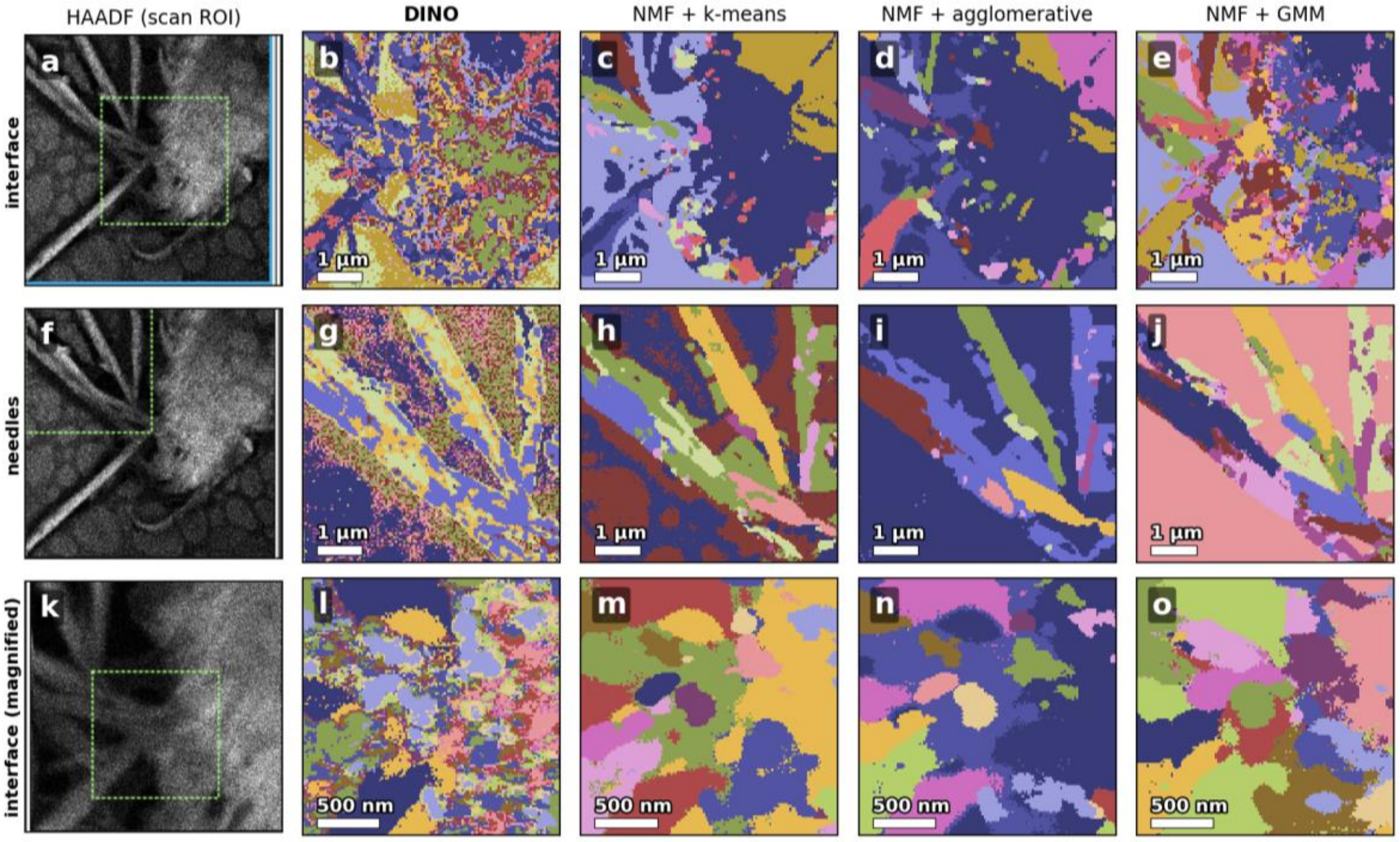

**Figure 4. Clustering of the indomethacin fields of view.** Each row is one region and each column one method. **a–e)** interface region: virtual HAADF with the 4D-STEM scan region outlined (a), the DINO4DSTEM class map (b), and polar NMF clustered by k-means (c), agglomerative (d) and Gaussian-mixture (e). **f–j)** needle region and **k–o)** magnified interface, in the same column order. Colors are arbitrary cluster labels. DINO4DSTEM, with the same settings throughout, returns one consistent map per region (b, g, l), whereas NMF, with parameters optimized for this sample, gives a different partition per algorithm. Agreement between NMF and DINO4DSTEM is low and algorithm-dependent (adjusted Rand index 0.05 to 0.27; 0 is chance-level and 1 an exact match), and the three NMF clusterings agree among themselves only at 0.15 to 0.69. Two further NMF clusterings (HDBSCAN and fuzzy-c-means) and the per-algorithm comparison are in Figure S10.

## 2.4. Evolving crystals of indomethacin

DINO4DSTEM was next applied to resolve the nanoscale structural evolution during indomethacin crystallization (Fig. 5). Because the framework segments the 4D-STEM dataset into structurally homogeneous nanodomains, diffraction patterns within each class can be averaged to recover high signal-to-noise diffraction signatures that are inaccessible from individual low-dose measurements. This enables crystallographic interpretation of each structural state while preserving its spatial distribution.

The class-average diffraction patterns reveal that crystallization proceeds through a continuous evolution of structural order rather than abrupt phase transitions. Across the ordering sequence, the dominant diffraction ring remains fixed at $d \approx 4.5–4.8$ Å, consistent with the α-indomethacin {102}/{112}[14] (Fig. S11), while its azimuthal distribution evolves continuously from an almost uniform diffuse ring to discrete Bragg reflections (Figs. 5 and S13). The least ordered nanodomains therefore already possess the characteristic intermolecular spacing of α-indomethacin but lack long-range orientational order (Fig. S12). Intermediate domains bridge these extremes, exhibiting measurable Bragg scattering despite lacking the characteristic needle morphology. Grain-resolved diffraction further shows that intact needle crystals index predominantly to α-indomethacin viewed close to the [100] zone axis, whereas interfacial domains remain substantially more diffuse (Fig. 5). Remarkably, even within the needle field, neighboring grains span nearly the full range of crystalline order, demonstrating that apparently mature needles remain structurally heterogeneous at the nanoscale.

To quantify the ordering trend, we computed three established, rotation-invariant diffraction descriptors for every DINO4DSTEM class: azimuthal spottiness[22], the two-dimensional Bragg excess ($B$), and the radial peak-to-halo ratio ($\chi$) (Experimental Section)[23]. Across all independently analyzed regions, each descriptor increases monotonically with the DINO4DSTEM class (Fig. 6), demonstrating that the learned classes define a single crystallinity axis extending continuously from amorphous to crystalline material. The resulting spatial maps reveal pronounced crystallinity gradients, with the highest order confined to the interiors of crystalline needles and progressively lower order extending into the surrounding matrix.

Across all three regions, the three independent, rotation-invariant diffraction descriptors increase monotonically with the DINO4DSTEM classes, defining a quantitative crystallinity axis that captures most of the descriptor variance ($\eta^2 = 0.72–0.98$; Fig. 6). The strong correlation between the order parameter and the DINO4DSTEM classes demonstrates that the learned representation encodes physically quantifiable structural properties. DINO4DSTEM thereby resolves a highly heterogeneous nanoscale structural landscape, revealing crystallinity gradients and coexisting intermediate nanodomains within and around individual crystalline needles. These quantitative maps show that molecular order develops continuously during crystal growth, a hallmark of the nonclassical crystallization mechanism[24–27].

This structural pathway is accessible because DINO4DSTEM first groups structurally related diffraction patterns before averaging them. Individual low-dose diffraction patterns contain insufficient signal to quantify crystallinity reliably, whereas class-wise averaging recovers robust diffraction signatures that expose the underlying ordering sequence. The framework therefore functions not only as a segmentation method but also as a means of extracting structural information hidden below the noise of individual measurements (Figs. 5-6). The same crystallization pathway was reproduced across multiple independent regions of the annealed films (Figs. S17–S19), demonstrating that it is a robust characteristic of indomethacin crystallization rather than a local feature.

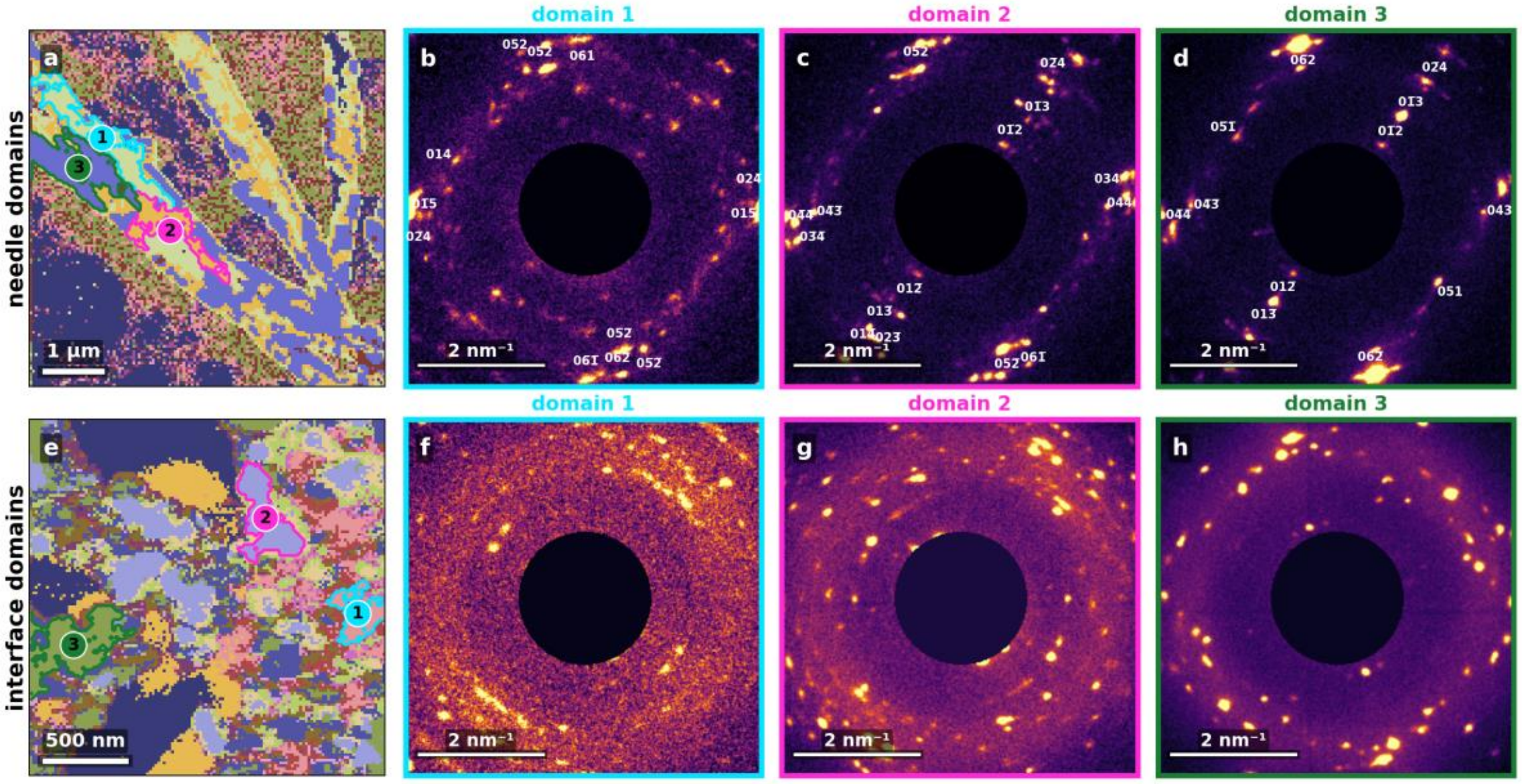

**Figure 5. Grain-resolved diffraction of the indomethacin nanodomains. a)** DINO4DSTEM class map of the needle field, coloured as in Figure 4, with three representative domains outlined and numbered 1 to 3 (scale bar, 1 µm). **b–d)** Grain-average diffraction of domains 1 to 3, ordered left to right by increasing crystalline order; the reflections index to α-indomethacin viewed close to the [100] zone axis, with representative hkl indices labelled. **e)** DINO4DSTEM class map of the interface field, with three representative domains outlined and numbered 1 to 3 (scale bar, 500 nm). **f–h)** Grain-average diffraction of those domains, ordered by increasing crystalline order; the patterns remain more diffuse, consistent with the less crystallized interface. Reciprocal scale bars in all diffraction panels, 2 $nm^{-1}$.

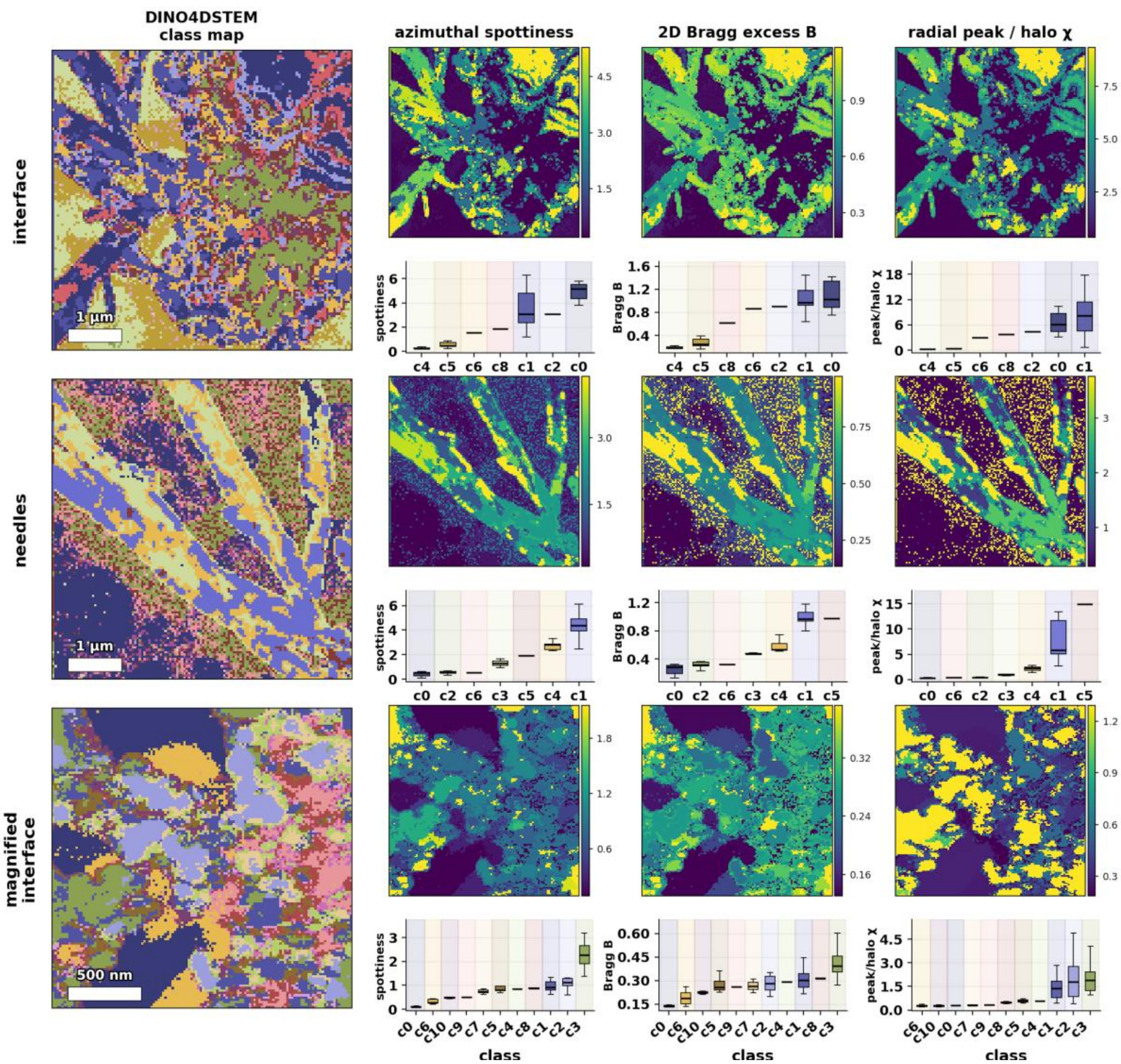

**Figure 6. Emergent DINO4DSTEM classes form a single crystallinity axis.** For each indomethacin region (rows: interface, needles, magnified interface): column 1, the DINO4DSTEM class map (coloured by class identity, as in Figure 5); columns 2 to 4, the three rotation-invariant descriptors (azimuthal spottiness, two-dimensional Bragg excess B, radial peak-to-halo ratio) shown as spatial maps (top; dark to bright is amorphous to crystalline) and as per-class distributions (bottom; each box is one class, classes ordered along the crystallinity axis, the annotated eta-squared is the fraction of the descriptor variance explained by the class label). The descriptors rise monotonically from class to class and peak along the needle morphology, identifying the classes as steps on a single amorphous-to-crystalline axis. The three descriptors agree on this ordering and on which classes are the most crystalline; they differ only in the ranking within that top group, where the fully crystalline classes are separated by the character of their order rather than its amount. Controls confirming that the descriptors are meaningful only after clustering are presented in Figure S20.

## 3. Conclusion

We have developed DINO4DSTEM, a self-supervised framework for quantitative structural state discovery in 4D-STEM that automatically identifies physically meaningful structural classes without structural models, predefined class numbers, or manual annotation. By grouping structurally related diffraction patterns directly from the data, the framework transforms large collections of weak, low-dose nanodiffraction measurements into quantitative nanoscale structural maps.

Applied to the challenging problem of crystallization in beam-sensitive indomethacin, DINO4DSTEM reveals that crystalline needles emerge from partially ordered precursor nanodomains rather than a fully amorphous state. It further shows that crystallization proceeds across a continuous spectrum of structural order, resolving a highly heterogeneous nanoscale landscape of crystallinity characterized by pronounced spatial gradients and coexisting intermediate states. By defining a quantitative crystallinity axis directly from the diffraction data, the framework provides new insights into crystal-growth pathways that are inaccessible from individual low-dose diffraction patterns.

DINO4DSTEM learns directly from the diffraction data which structural differences are meaningful, it automatically identifies the dominant structural degree of freedom in each dataset: crystallinity in indomethacin, diffuse line-scattering domains and measurement-induced interfaces in NaPHI, and crystallographic viewing direction in $WS_2$. In each case, the relevant structure axis emerges without prior knowledge or dataset-specific tuning. DINO4DSTEM therefore establishes a general framework for quantitative nanoscale structural mapping from 4D-STEM, enabling automated discovery, visualization, and quantification of structural organization in complex, heterogeneous, and beam-sensitive materials.

## 4. Experimental Section

**Sample preparation:** Indomethacin films (approximately 150 nm) were grown by physical vapour deposition at 0.2 Å $s^{-1}$ onto lacy amorphous-carbon-coated gold TEM grids, forming an amorphous film, and then annealed ex situ at 70 °C for 60 min under nitrogen, above the glass transition (about 45 °C) and below the 155–162 °C melt, to nucleate and grow crystalline α-needles. Polarized-light microscopy showed nucleation starting at the lacy-carbon edges, consistent with strain induced on the film. The sodium poly(heptazine imide) (NaPHI) flake and its reference SAM segmentation are from prior work.[12]

**4D-STEM acquisition:** Energy-filtered nanobeam 4D-STEM was obtained on a double corrrected Thermo Fisher Scientific Themis-Z scanning electron microscope, equipped with a CEOS CEFID energy filter, at 200 kV electron energy, a beam current of 3 to 4 pA, an approximately 20 nm probe, and with a 16 eV energy-selecting slit. Diffraction frames were collected with a DECTRIS ELA detector in electron counting mode with a reciprocal space sampling of 0.00185 $Å^{-1}$ per pixel. Each dataset is a 128 × 128 probe-position scan. The interface and needle regions were acquired at 7,100× magnification with a 44 nm step (about 5.7 µm field) and 5 ms per pattern (dose about 50 $e^{-}$ $Å^{-2}$); the same needle/matrix interface at higher magnification was acquired at 20,000× with a 16 nm step (about 2.0 µm field) and 10 ms per pattern (dose about 100 $e^{-}$ $Å^{-2}$). Series of repeated frame aquisitions proved that the doses were low enough to avoid beam-induced crystallization.

**Simulated benchmark sample:** To obtain a ground-truthed test of the classifier, we generated a simulated polycrystalline WS2 (2H) sample with known orientation labels and computed its 4D-STEM diffraction by multislice simulation. The atomic model was a WS2 (2H) layered structure (lattice parameter a = 3.153 Å) built with the ASE mx2 builder and tiled into a 14 by 14 by 6 orthogonal supercell, six layers thick. The microstructure was a Voronoi tessellation of 64 grains over the 96 by 96 probe field; each grain was assigned at random one of three crystallographic zone-axis orientations, the basal [001] and the two edge-on [110] and [100] directions, together with an independent in-plane rotation drawn uniformly between 0 and 360 degrees. This defines three ground-truth classes, one per zone axis, with the in-plane rotation acting as a nuisance variable that a physically meaningful, rotation-invariant classifier must ignore. Simulated diffraction was computed with abTEM using the finite-projection multislice engine at 200 kV, with a 1.5 mrad convergence semi-angle, on a 256 by 256 pixel detector spanning 28 mrad, at 96 by 96 probe positions on a 3.0 Å step, giving a real-space sampling of 3.0 Å per scan pixel and a reciprocal-space sampling of 0.0872 nm per detector pixel; an idealized high dose was used so that the simulated patterns are effectively noiseless (random seed 0). The identical DINO4DSTEM classifier and training recipe used for the experimental data were then applied to this simulated sample for 30 epochs without labels, and the predicted classes were matched to the ground-truth orientations by optimal (Hungarian) assignment (Figures S3 to S6).

**Model and training:** DINO4DSTEM uses self-distillation[11] with a student encoder and a slowly updated (exponential moving-average) teacher encoder. The encoder is a ResNet-18[18] truncated to its first residual stage (layer-1, two BasicBlocks), with deeper stages selectable up to four. The main augmentation is a random in-plane rotation that enforces orientation invariance, applied together with light additional augmentation (Gaussian blur and central-beam masking); a one-dimensional radial clustering loss requires each pattern's azimuthally averaged profile to match its assigned class and to differ from the others. An optional semi-supervised pair loss is available. The same hyperparameters were used for every dataset in this study: 60 prototypes, 30 epochs, batch size 128, learning rate $3 \times 10^{-4}$, center momentum m = 0.97, teacher EMA from 0.99 to 0.999, and a polar input representation. Only per-sample pre-processing was adjusted (centering, central-beam mask radius, and intensity scaling). The choice of encoder, its depth, and the loss terms are examined in the Supporting Information (Figures S1 and S2).

**Crystallinity descriptors:** Three rotation-invariant descriptors were computed from the average diffraction pattern of each class and each grain, taken as the unmasked mean of all scan positions assigned to it. In each pattern the direct beam was masked to a radius of eleven percent of the pattern width, and the analysis was confined to a radial window reaching from just outside that mask to a scattering vector of 0.35 $Å^{-1}$, or to the edge of the recorded field where that was smaller. Within this window we computed, in one-pixel radial bins, the azimuthal mean and the azimuthal variance of intensity, and estimated the smooth amorphous halo with a SNIP iterative peak-clipping baseline (fourteen passes of growing half-window) applied to the logarithm of the azimuthal mean and then exponentiated; the peak component is the azimuthal mean minus this halo, clipped at zero. (i) The azimuthal spottiness is the ninetieth percentile, over the radial window, of the azimuthal coefficient of variation, that is the standard deviation of intensity around each ring divided by its mean: a smooth amorphous ring is nearly uniform and gives a value close to zero, whereas discrete Bragg reflections produce bright maxima and dark gaps around the ring and a large value. (ii) The two-dimensional Bragg excess B is a full-pattern measure: the one-dimensional halo is interpolated onto every pixel to form a two-dimensional background, and B is the intensity summed above this background over the annular window divided by the summed background, so it captures scattered intensity rising above the diffuse halo anywhere in the pattern rather than only on the azimuthal average. (iii) The radial peak-to-halo ratio $\chi$ is the maximum over the window of the peak component divided by the halo, the height of the strongest ring above the local background relative to that background; it is near zero for a pure halo and large when a sharp reflection dominates. All three are rotation-invariant, being built from azimuthal averages and full-pattern sums with no angular reference, and are reported per class and per grain rather than per frame, where shot noise dominates (Figure S20).

**Validation of the precursor assignment:** Three checks support interpreting the least-ordered on-sample region as a genuine, partly ordered state. Its average diffraction is consistent with the α-indomethacin structure, with the principal halo at the {102}/{112} reflections near d ≈ 4.75 Å and a weaker feature at {103} (3.90 Å), at the same d-spacings as the mature needles but azimuthally diffuse. The film is amorphous as deposited, as established for vapour-deposited indomethacin[20,28] and as is evident from previous experiments (Supporting Information); the specific polymorph is not central to this analysis, which concerns the degree of order rather than the phase, and the powder-averaged radial profiles do not by themselves distinguish the α and γ polymorphs (Supporting Information). The region is thick and high-scattering (above-median scattered intensity) rather than thin or empty, so its single frames are shot-noise limited. And the imaging dose (about 50 $e^- Å^{-2}$ for the interface and needle regions, 100 $e^- Å^{-2}$ for the interface) is well below that associated with beam-induced crystallization, while radiation damage would reduce order rather than create it.

**Comparison methods and indexing:** Non-negative matrix factorization[13] was computed on a rotation-invariant polar representation and clustered by k-means, agglomerative, Gaussian-mixture, HDBSCAN,[19] and fuzzy-c-means; agreement with the DINO4DSTEM maps is reported as the adjusted Rand index. This index measures how much two label maps agree beyond chance: it is 0 for independent partitions and 1 for identical ones, and is corrected for the number and size of the clusters.[15] The α-indomethacin reflections were assigned from the α crystal structure[5] by computed kinematical intensity (the {102}/{112} pair at d ≈ 4.75–4.83 Å and {103} at 3.90 Å, with {022} and {003} at larger d).

**Software, code, and data availability:** The complete pipeline is released as open-source software at github.com/DanielKhaykelson/dino-4dstem. It comprises the DINO4DSTEM model with pre-tuned parameters, a graphical interface, and a natural-language assistant. The graphical interface lets a non-specialist load a dataset, run the fixed-configuration model, and inspect the result at three scales, a single diffraction frame, a grain, or a whole class, with gradient-based attribution (Grad-CAM)[20] indicating which part of a pattern drove each class assignment, and offers several merging protocols to combine classes without retraining. The natural-language assistant drives the same workflow, from loading and training through clustering, inspection, and export, from plain-language requests, so no coding is required. Screenshots and a walkthrough are in Figure S14. The 4D-STEM datasets are available from the corresponding author on request.

**References:**

1. Bernstein, J. Polymorphism and structure-property relations in molecular crystals. *Acta Crystallogr. A* **62**, s110–s110 (2006).
2. Davey, R. J. Polymorphism in molecular crystals Joel Bernstein. Oxford university press, New York, 2002. ISBN 0198506058. *Cryst. Growth Des.* **2**, 675–676 (2002).
3. Du, J. S., Bae, Y. & De Yoreo, J. J. Non-classical crystallization in soft and organic materials. *Nat. Rev. Mater.* **9**, 229–248 (2024).
4. Tsarfati, Y. *et al.* Crystallization of organic molecules: Nonclassical mechanism revealed by direct imaging. *ACS Cent. Sci.* **4**, 1031–1036 (2018).
5. Ophus, C. Four-dimensional scanning transmission electron microscopy (4D-STEM): From scanning nanodiffraction to ptychography and beyond. *Microsc. Microanal.* **25**, 563–582 (2019).
6. Bustillo, K. C. *et al.* 4D-STEM of Beam-Sensitive Materials. *Acc. Chem. Res.* **54**, 2543–2551 (2021).
7. Ophus, C. *et al.* Automated crystal orientation mapping in py4DSTEM using sparse correlation matching. *Microsc. Microanal.* **28**, 1–14 (2022).
8. Yoo, T. *et al.* Unsupervised machine learning and cepstral analysis with 4D-STEM for characterizing complex microstructures of metallic alloys. *Npj Comput. Mater.* **10**, 1–10 (2024).
9. Kirillov, A. *et al.* Segment Anything. *arXiv [cs.CV]* (2023) doi:10.48550/arXiv.2304.02643.

10. Lee, S. *et al.* Unsupervised segmentation and clustering workflow for efficient processing of 4D-STEM and 5D-STEM data. *arXiv [cond-mat.mtrl-sci]* (2026) doi:10.48550/arXiv.2601.17262.

11. Cao, J. *et al.* Unsupervised multi-clustering and decision-making strategies for 4D-STEM orientation mapping. *Digit. Discov.* **4**, 3610–3622 (2025).

12. Kimoto, K. *et al.* Unsupervised machine learning combined with 4D scanning transmission electron microscopy for bimodal nanostructural analysis. *Sci. Rep.* **14**, 2901 (2024).

13. Bridger, A., David, W. I. F., Wood, T. J., Danaie, M. & Butler, K. T. Versatile domain mapping of Scanning Electron Nanobeam Diffraction datasets utilising variational AutoEncoders and decoder-assisted latent-space clustering. *arXiv [cond-mat.mtrl-sci]* (2022) doi:10.48550/arXiv.2207.13389.

14. Stowell, J. G., Chen, X., Morris, K. R., Griesser, U. J. & Byrn, S. R. Reactivity differences of indomethacin solid forms with ammonia gas. *Acta Crystallogr. A* **58**, c143–c143 (2002).

15. Cox, J. R., Ferris, L. A. & Thalladi, V. R. Selective growth of a stable drug polymorph by suppressing the nucleation of corresponding metastable polymorphs. *Angew. Chem. Weinheim Bergstr. Ger.* **119**, 4411–4414 (2007).

16. Andrusenko, I. *et al.* Structure determination, thermal stability and dissolution rate of δ-indomethacin. *Int. J. Pharm.* **608**, 121067 (2021).

17. Cañellas, F. M. *et al.* Controlling the polymorphism of indomethacin with poloxamer 407 in a gas antisolvent crystallization process. *ACS Omega* **7**, 43945–43957 (2022).

18. Caron, M. *et al.* Emerging properties in self-supervised vision transformers. *arXiv [cs.CV]* (2021).

19. Khaykelson, D. *et al.* Elucidating structural disorder in a polymeric layered material: The case of sodium poly(heptazine imide) photocatalyst. *Nano Lett.* **25**, 17230–17236 (2025).

20. Dawson, K. J., Kearns, K. L., Yu, L., Steffen, W. & Ediger, M. D. Physical vapor deposition as a route to hidden amorphous states. *Proc. Natl. Acad. Sci. U. S. A.* **106**, 15165–15170 (2009).

21. Tran, V. A. *et al.* D-IMPACT: A data preprocessing algorithm to improve the performance of clustering. *J. Softw. Eng. Appl.* **07**, 639–654 (2014).

22. Treacy, M. M. J., Gibson, J. M., Fan, L., Paterson, D. J. & McNulty, I. Fluctuation microscopy: a probe of medium range order. *Rep. Prog. Phys.* **68**, 2899 (2005).

23. Donohue, J. *et al.* Cryogenic 4D-STEM analysis of an amorphous-crystalline polymer blend: Combined nanocrystalline and amorphous phase mapping. *iScience* **25**, 103882 (2022).

24. Sedova, A., Houben, L., Seifer, S., Cohen, S. R. & Rybtchinski, B. Evolution of multi-lamellar crystals in thermoplastic revealed by 2D and 3D nanodiffraction imaging. *Commun. Mater.* **7**, (2026).

25. Biran, I. *et al.* Real-space crystal structure analysis by low-dose focal-series TEM imaging of organic materials with near-atomic resolution. *Adv. Mater.* **34**, e2202088 (2022).

26. Tsarfati, Y. *et al.* Continuum crystallization model derived from pharmaceutical crystallization mechanisms. *ACS Cent. Sci.* **7**, 900–908 (2021).

27. Biran, I. *et al.* Organic crystal growth: Hierarchical self-assembly involving nonclassical and classical steps. *Cryst. Growth Des.* **22**, 6647–6655 (2022).

28. Dawson, K. J., Kearns, K. L., Ediger, M. D., Sacchetti, M. J. & Zografi, G. D. Highly stable indomethacin glasses resist uptake of water vapor. *J. Phys. Chem. B* **113**, 2422–2427 (2009).

# Supporting Information

A central aim of this work is to bridge the gap between electron microscopy and materials science by making advanced structural analysis accessible to non-specialists. Because the network operates with fixed settings across all datasets, user input is limited to physically motivated preprocessing, enabling reproducible mapping across samples and operators without sample-specific tuning. The complete pipeline, including a graphical interface for multi-scale inspection with gradient-based attribution (Grad-CAM)[20] and a natural-language assistant that runs the workflow from plain-language requests, is released as open-source software (see the Experimental Section, Figure S14, and github.com/DanielKhaykelson/dino-4dstem).

This Supporting Information contains the architecture and loss ablations, the per-class composition of the NaPHI segmentation, the clustering-dependence of NMF, the controls and reproducibility tests behind the indomethacin crystallinity analysis, and the introductory graphical user interface(GUI) description and manuals. Every DINO4DSTEM map uses the same training settings as the main results; only the component under test, or the dataset, is varied.

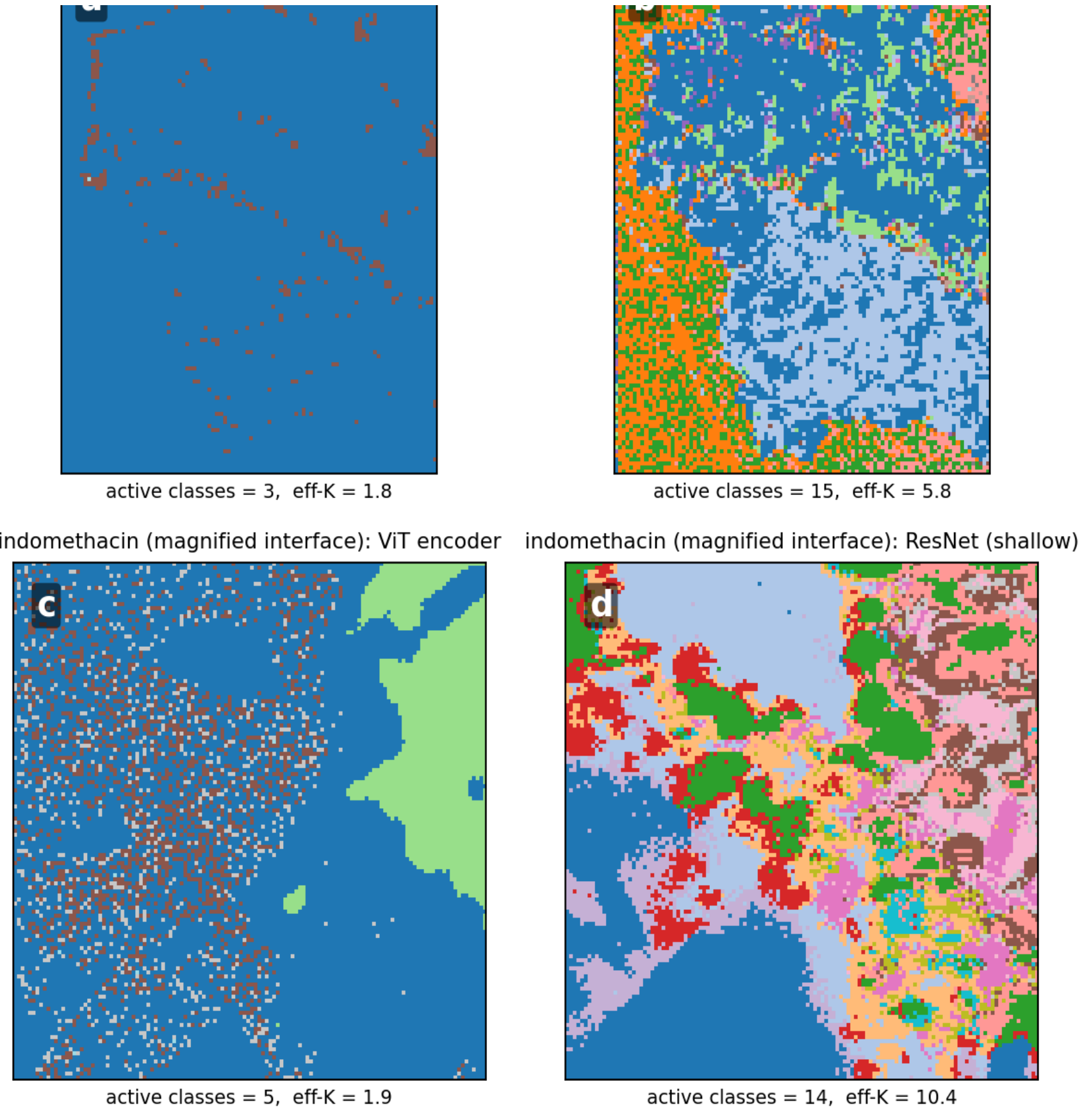

**Figure S1. Encoder architecture.** Class maps for NaPHI (Na007b) and indomethacin (magnified interface) using a vision-transformer encoder (a, c) versus the shallow convolutional encoder at matched depth (b, d). The transformer collapses to a few classes, while the convolutional encoder recovers the full domain structure. Each panel is annotated with the number of active classes and the effective number of classes.

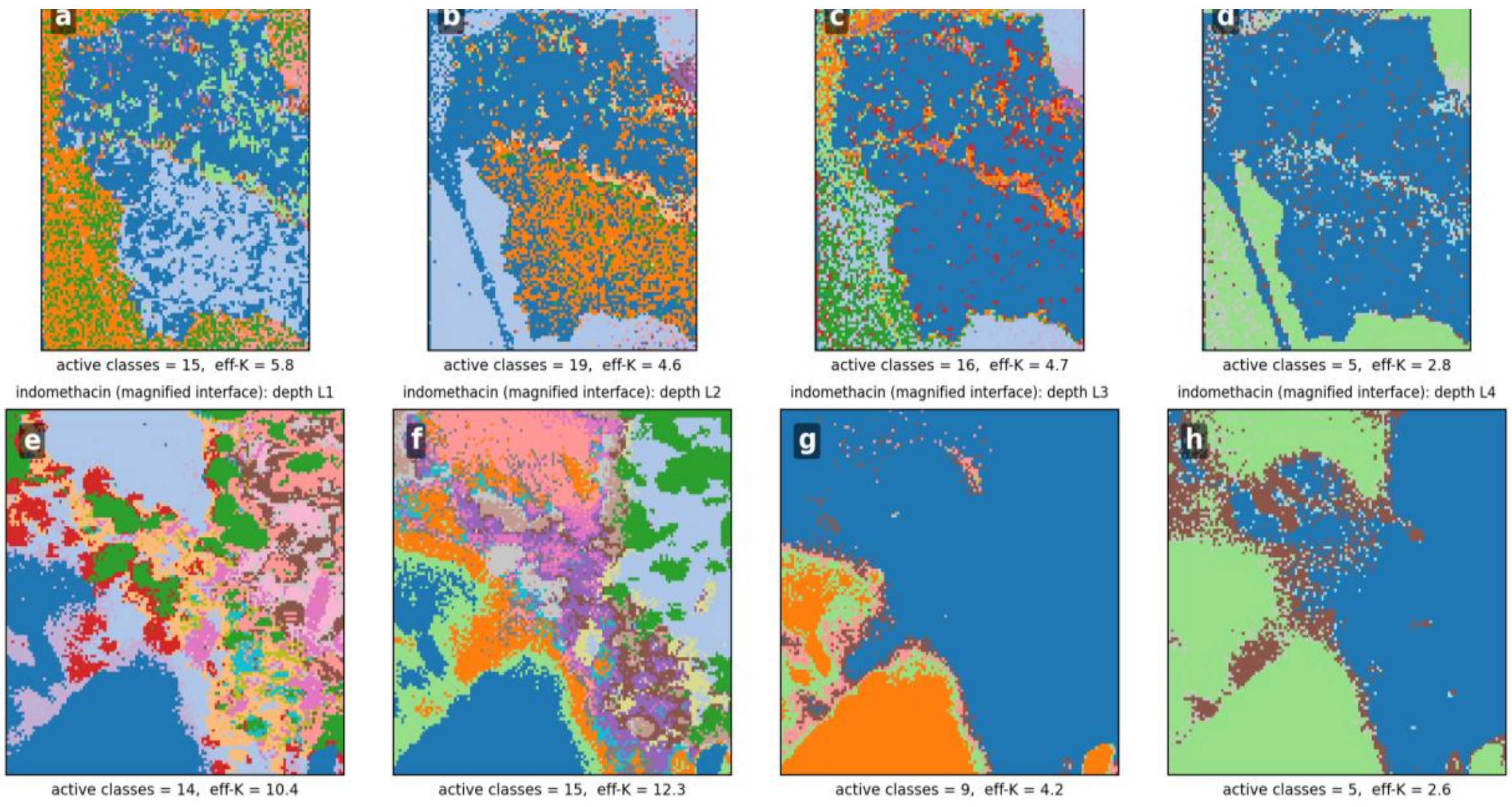

**Figure S2. Encoder depth.** Class maps as the convolutional encoder is deepened from its first residual stage (L1, two BasicBlocks) to four (L4), for NaPHI (a to d) and indomethacin (e to h). Shallow encoders (L1, L2) retain the domain structure; deeper encoders (L3, L4) collapse to fewer, coarser classes, which is why a shallow trunk is used throughout.

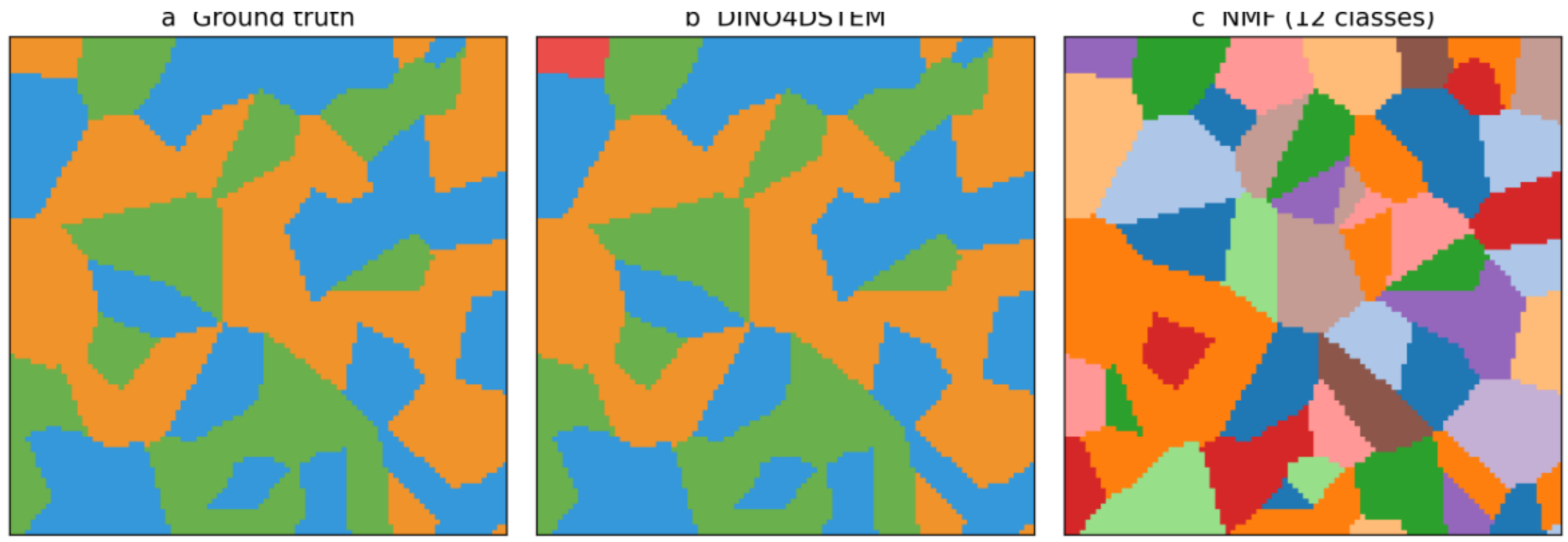

**Figure S3.** Synthetic validation on a ground-truthed WS2 (2H) polycrystalline simulated sample (class maps). a) Ground-truth orientation map of the 64-grain microstructure, coloured by the three assigned three-dimensional zone-axis orientations ([001], [110], [100]). b) DINO4DSTEM class map (unsupervised, no labels), with cluster colours matched to the ground truth by Hungarian assignment; the map is visually near-identical to (a), with grains of the same orientation but different in-plane rotation grouped into one class (pixel accuracy 98.9%, adjusted Rand index 0.982, normalized mutual information 0.978; per-class F1 = 0.985, 1.000, 1.000). c) Polar NMF class map of the same simulated sample, which over-segments the microstructure into many more classes than the three physical orientations, splitting each orientation by its in-plane rotation into approximately one class per grain (twelve classes at the tested cluster-number cap).

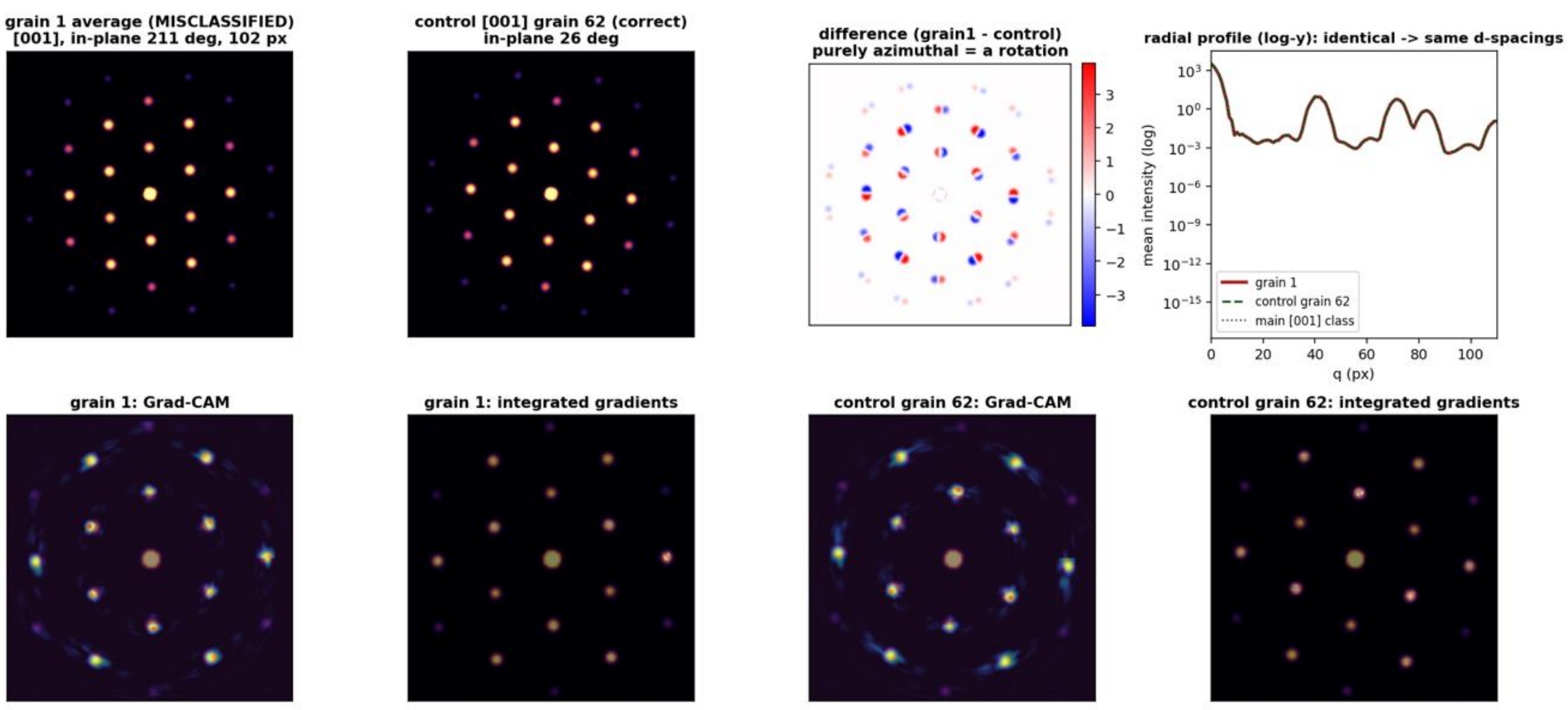

**Figure S4.** Explaining the single misclassified grain with the framework's own analysis tools. Top: the grain's class-average diffraction versus a correctly-classified [001] control grain, their difference (a purely azimuthal dipole at each spot, i.e. an in-plane rotation), and their radial profiles (log-scale y, identical, confirming the same d-spacings). Bottom: Grad-CAM and integrated gradients for both grains, evaluated against the [001] prototype, attend to the same Bragg reflections. The grain is genuinely [001]; it detached only because it is the single most atypical [001] grain in the scan, sitting at the assignment boundary between two near-identical prototypes.

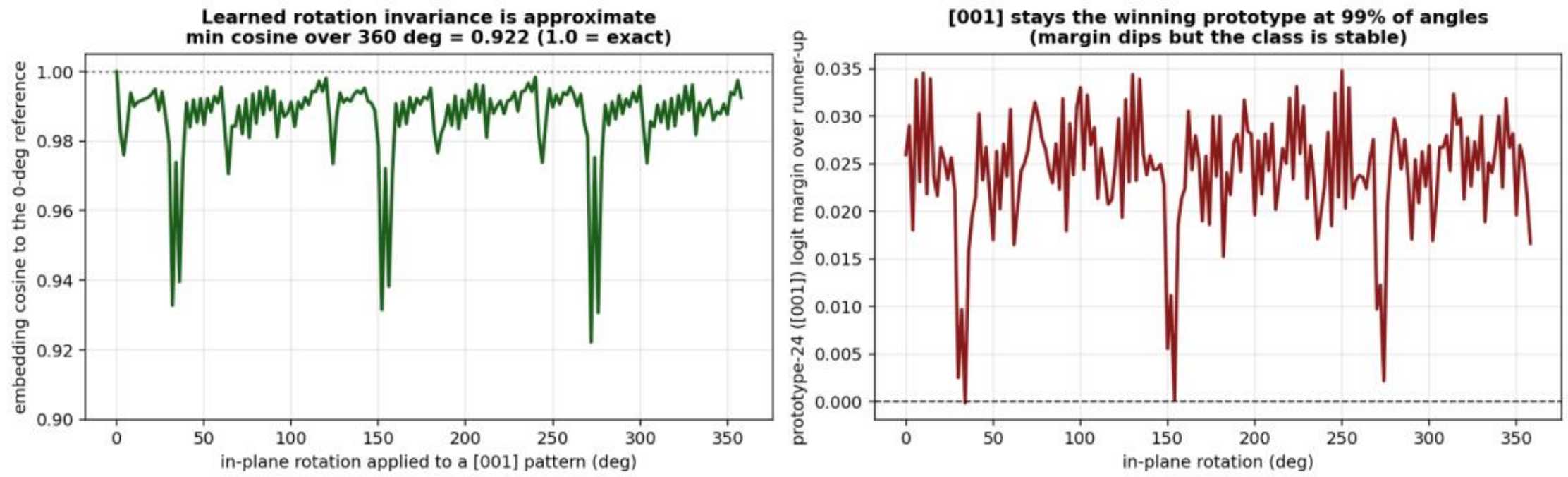

**Figure S5.** Rotation invariance is learned and therefore approximate, not exact. Left: sweeping a [001] pattern through all in-plane rotations, its 128-dimensional embedding stays at cosine 0.98 to 0.99 with the 0-degree reference over most of the range but dips to about 0.92 at a few symmetry-related angles (a perfectly invariant model would be flat at 1.0). Right: despite that embedding drift, the [001] prototype remains the winning class at 99.4% of in-plane angles, with the winning margin collapsing toward zero only at those same angles, which is where a borderline grain can tip onto a redundant prototype.

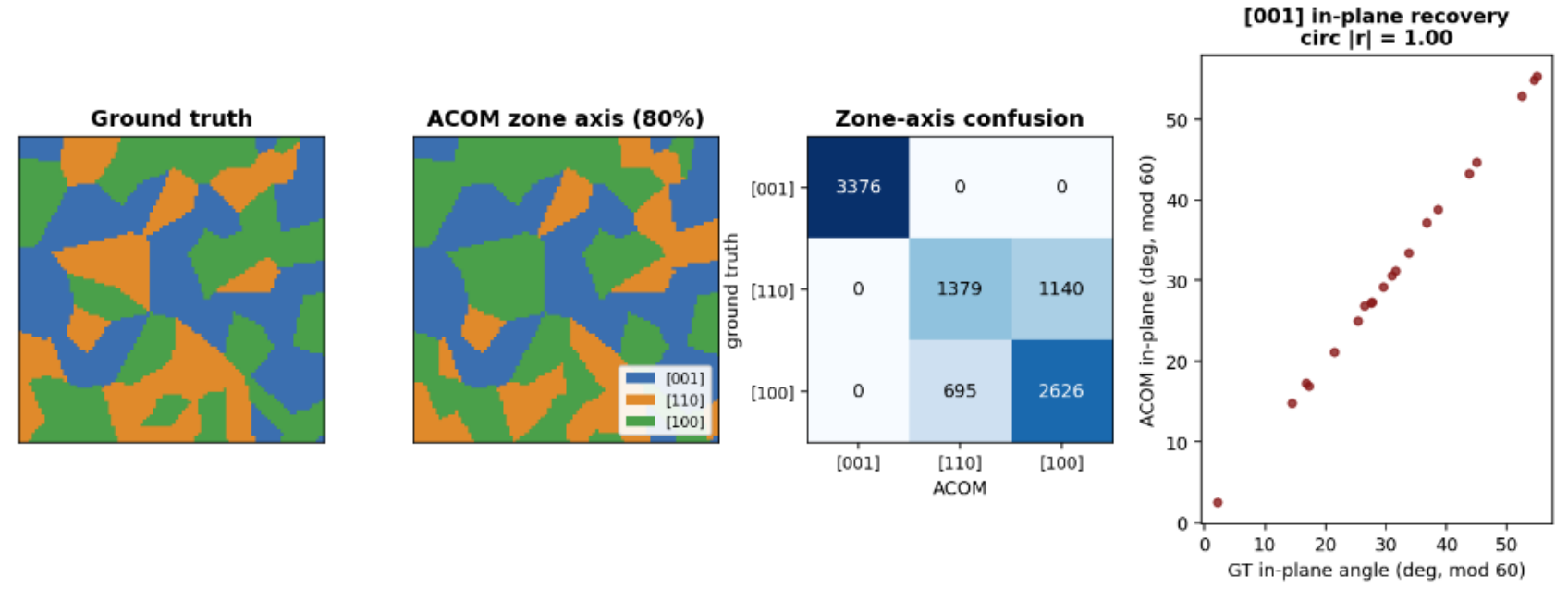

**Figure S6.** WS2 ACOM validated against ground truth. Left: the ground-truth phase / zone-axis map. Middle-left: the ACOM zone-axis map (80% pixel accuracy). Middle-right: the zone-axis confusion matrix ([001] recovered perfectly; [110] and [100] partly confused). Right: for the [001] grains, the ACOM-recovered in-plane rotation versus the ground-truth grain angle (modulo the 60-degree hexagonal symmetry), circular correlation 0.999.

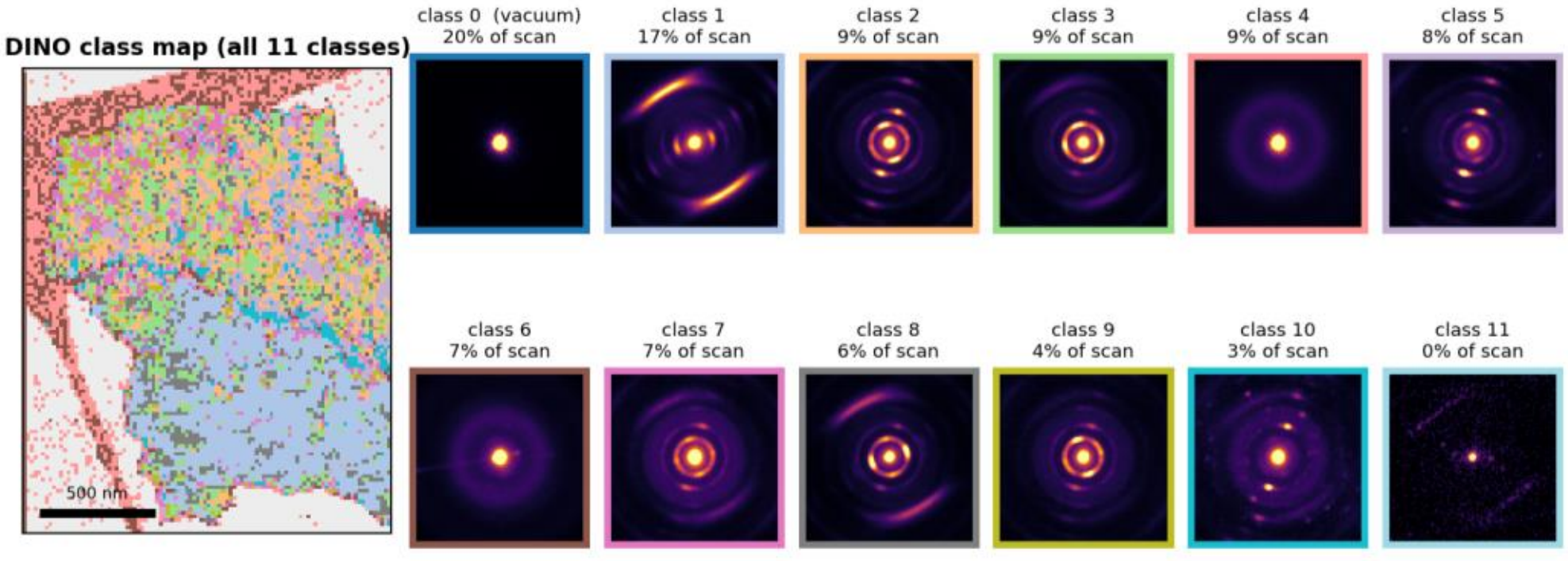

---

**Figure S7. NaPHI DINO4DSTEM classes and their meaning.** The full set of DINO4DSTEM classes for the NaPHI flake (class map at left) with the average diffraction pattern of each class. The line domain, the thickness sub-domains, and the measurement-induced interface each have a distinct average pattern, confirming that the extra classes in Figure S7g are genuine structural differences rather than over-segmentation.

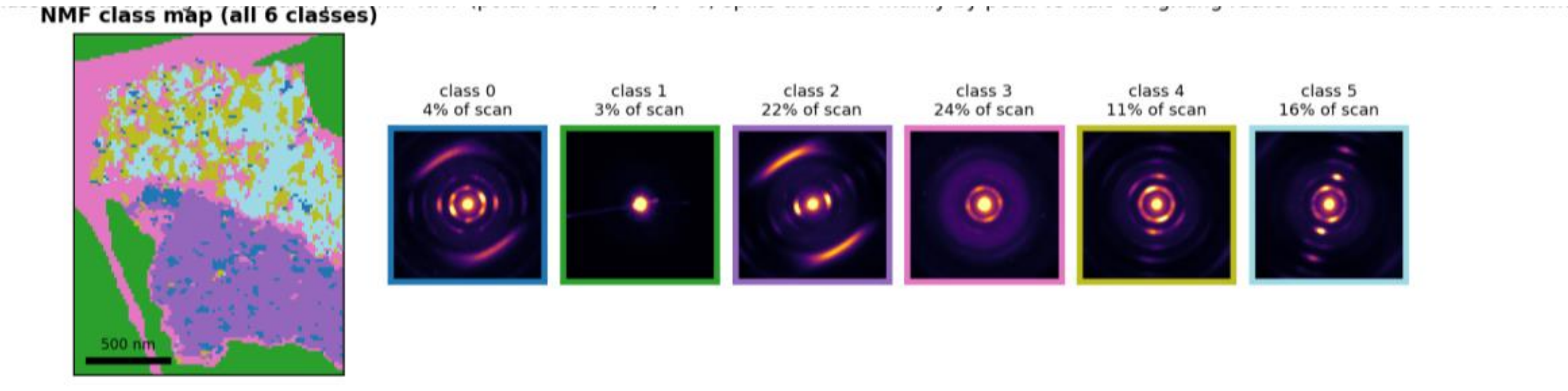

**Figure S8. NaPHI NMF classes.** The corresponding NMF (K = 6, polar with θ-shift) class map and per-class average diffraction, clustered identically to Figure 2. NMF separates the broad regions but does not resolve the finer thickness and interface sub-domains that DINO4DSTEM recovers.

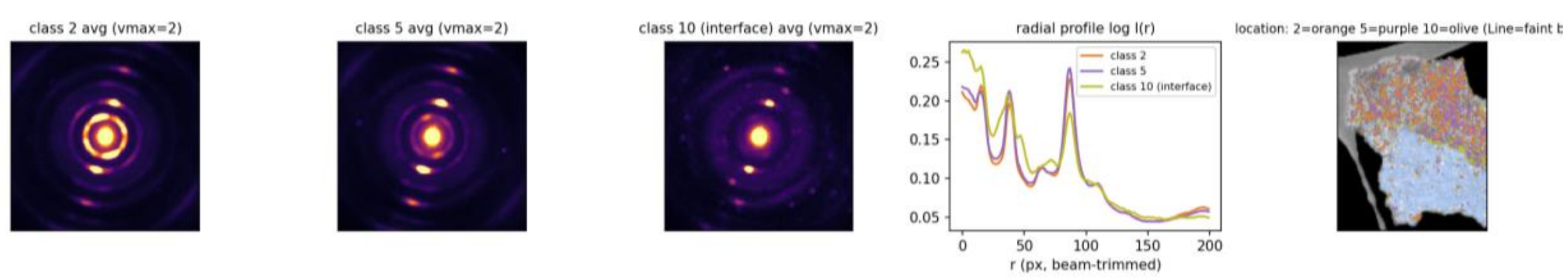

**Figure S9. The measurement-induced interface domain in NaPHI.** The interface class and its neighbours at the boundary between the line domain and the rest of the flake, with average diffraction, radial profiles, and spatial location. About 69% of its pixels border the line region, identifying it as a thin transition rather than a separate phase.

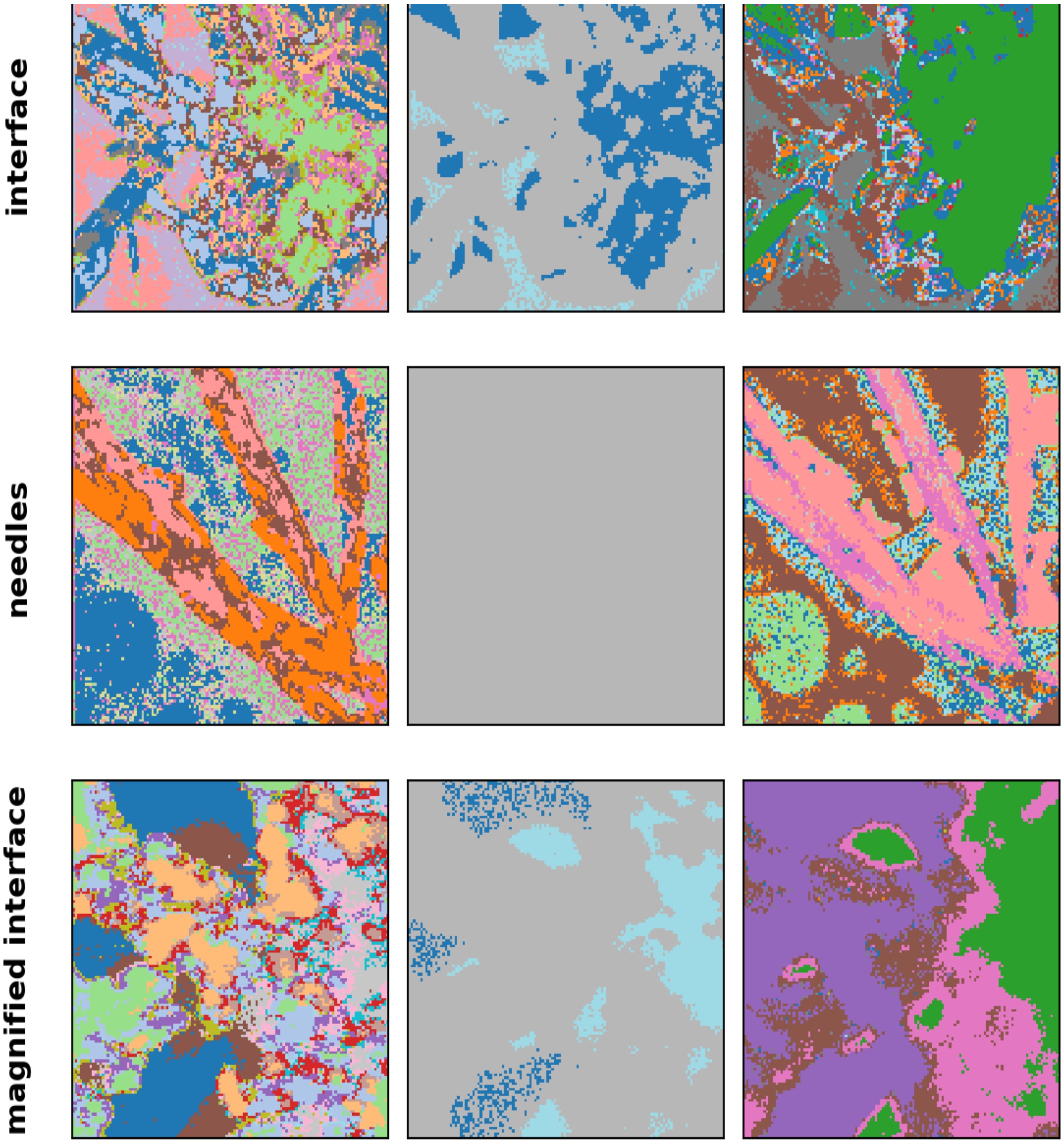

**Figure S10. NMF under further clustering algorithms.** The two NMF clusterings held out of Figure 4, HDBSCAN and fuzzy-c-means, for the three indomethacin regions, beside the DINO4DSTEM map. HDBSCAN collapses much of the field to noise and fuzzy-c-means gives yet another partition, so the NMF result depends strongly on the clustering algorithm, whereas the DINO4DSTEM map is fixed.

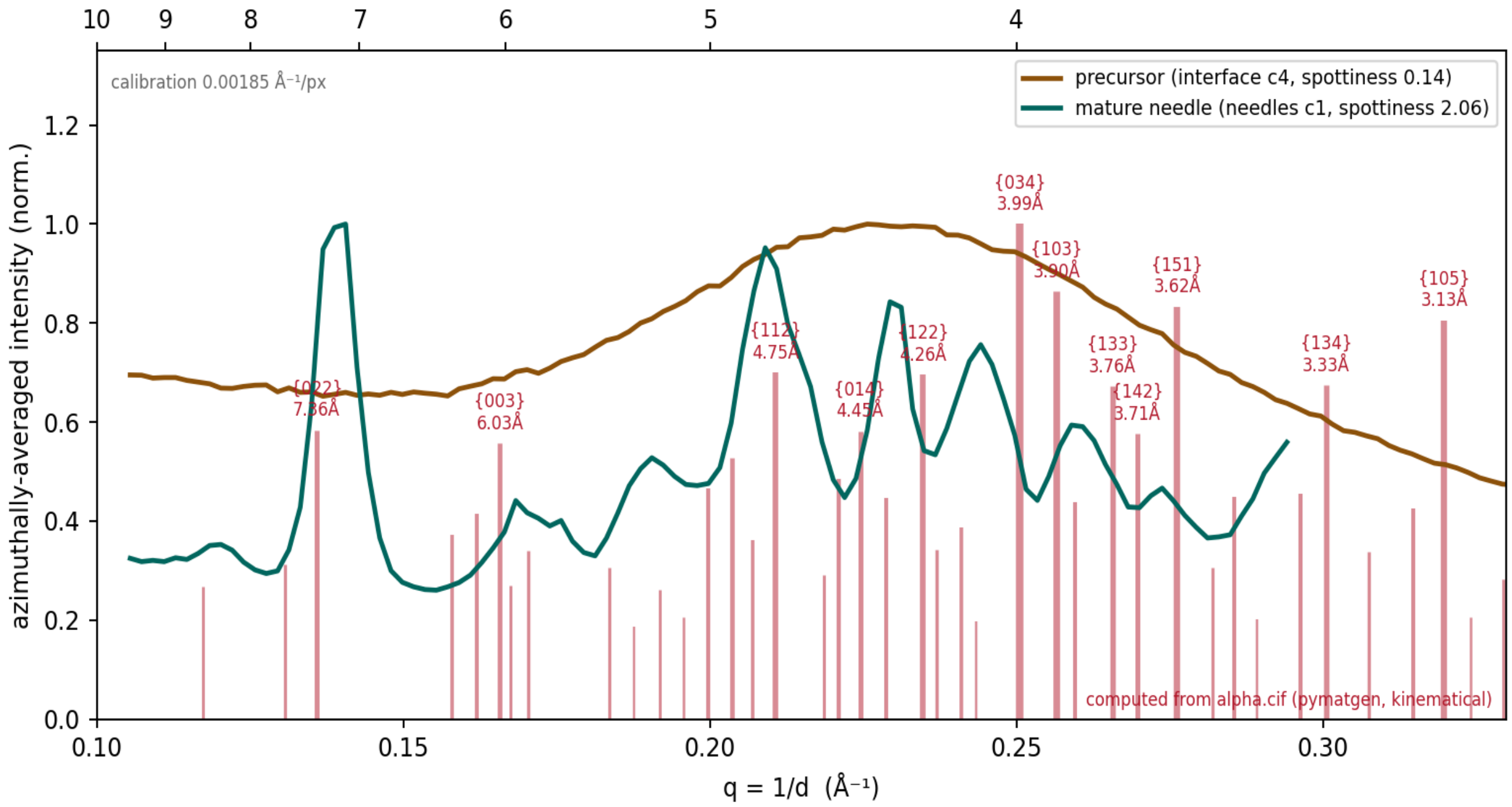

**Figure S11. Indexing of the precursor and needle, consistent with α-indomethacin.** Azimuthally averaged radial profiles of the least-ordered precursor and a mature needle on a reciprocal-space axis (q = 1/d; calibration 0.00185 $Å^{-1}$ per detector pixel), with the α-indomethacin reflections computed from the published crystal structure (alpha.cif; pymatgen, kinematical) overlaid as intensity-scaled sticks. The mature-needle peaks fall on the calculated α reflections across the whole range ({022} at 7.4 Å through {105} at 3.1 Å), and the precursor halo sits under the same envelope, peaking at the principal {102}/{112} ring (4.75 to 4.83 Å). This confirms that the precursor is partly ordered α-indomethacin rather than a distinct phase, and that the reciprocal-space calibration is consistent with the structure.

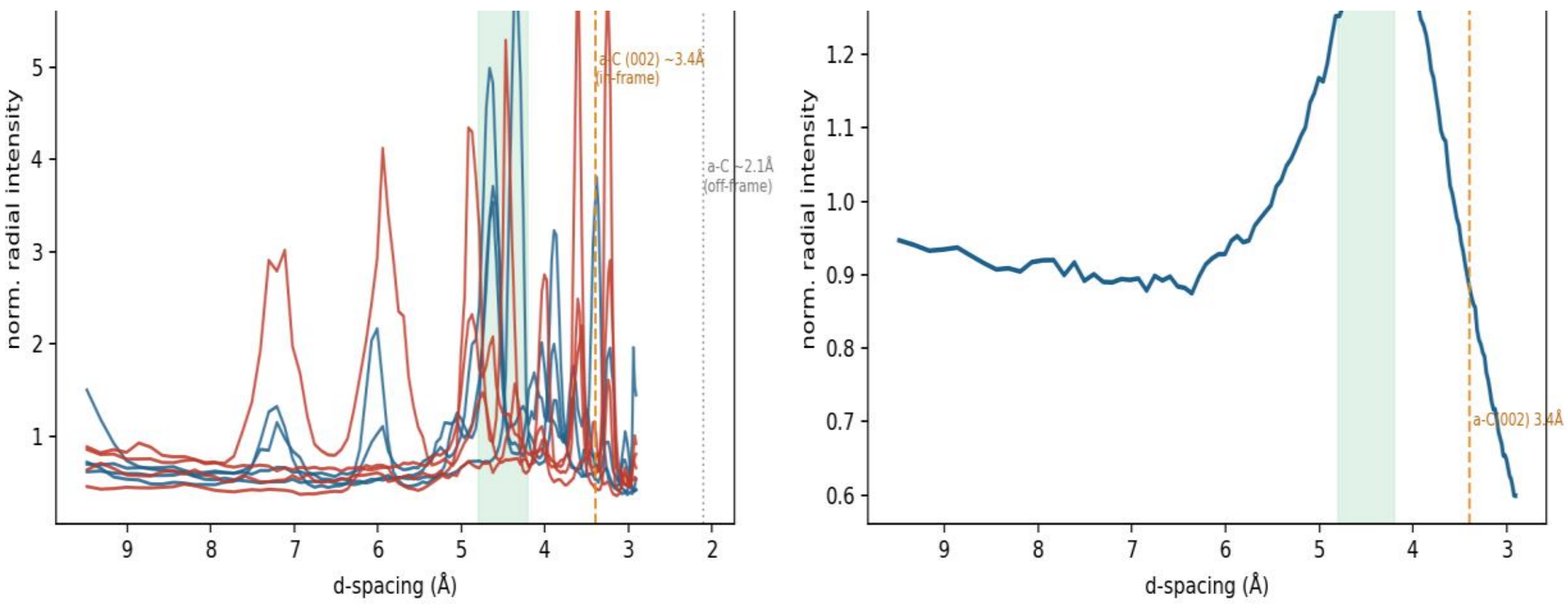

**Figure S12. Distinguishing the amorphous-carbon support from indomethacin.** A diffraction-only discriminator. Amorphous carbon's main halo lies at d ≈ 2.1 Å, outside the recorded window, so bare support is nearly featureless in our frame, whereas a clear d ≈ 4.5 Å ring is an organic (indomethacin) signature. Radial profiles of the thickest and thinnest interface grains show that the least-ordered matrix region carries the 4.5 Å ring and is therefore indomethacin, not carbon.

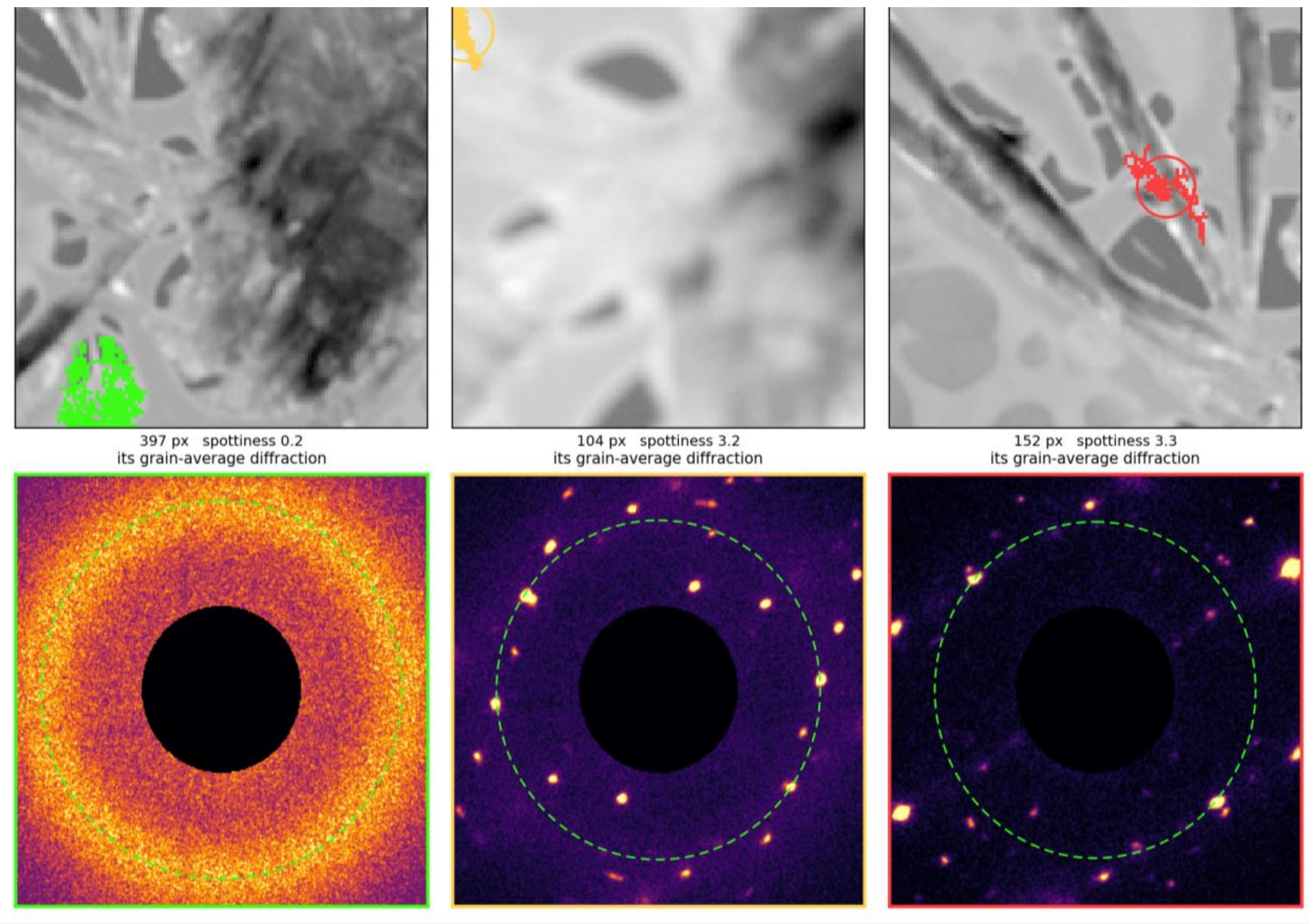

**Figure S13. Provenance of the representative diffraction patterns.** For each of the three representative grains used to illustrate the order axis (the less-ordered matrix, the crystallization front, and the mature needle), the selected grain's footprint is marked on the scattered-intensity map of the scan, with its grain-average diffraction beneath, so each representative pattern is traceable to a location on the sample.

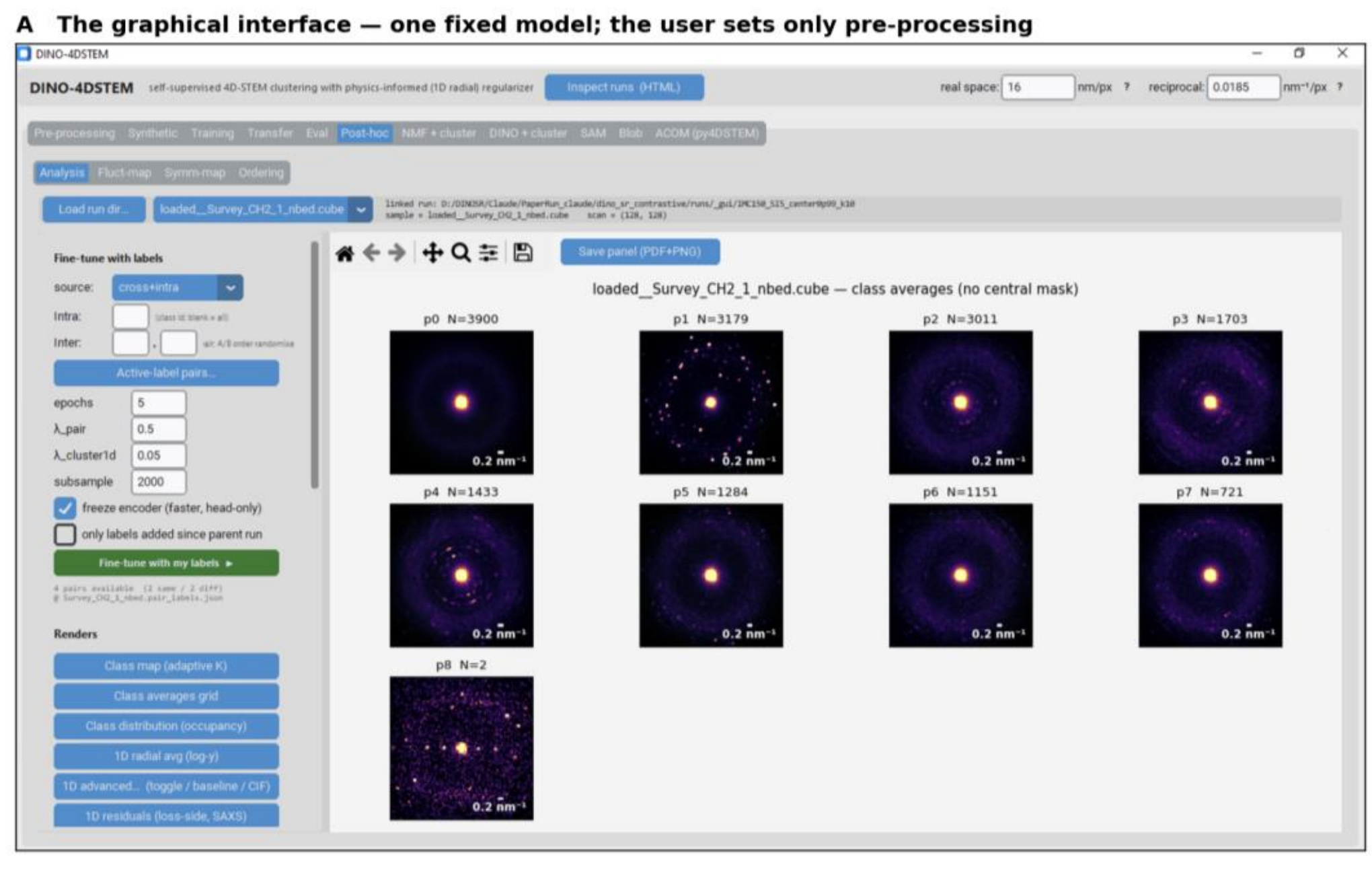

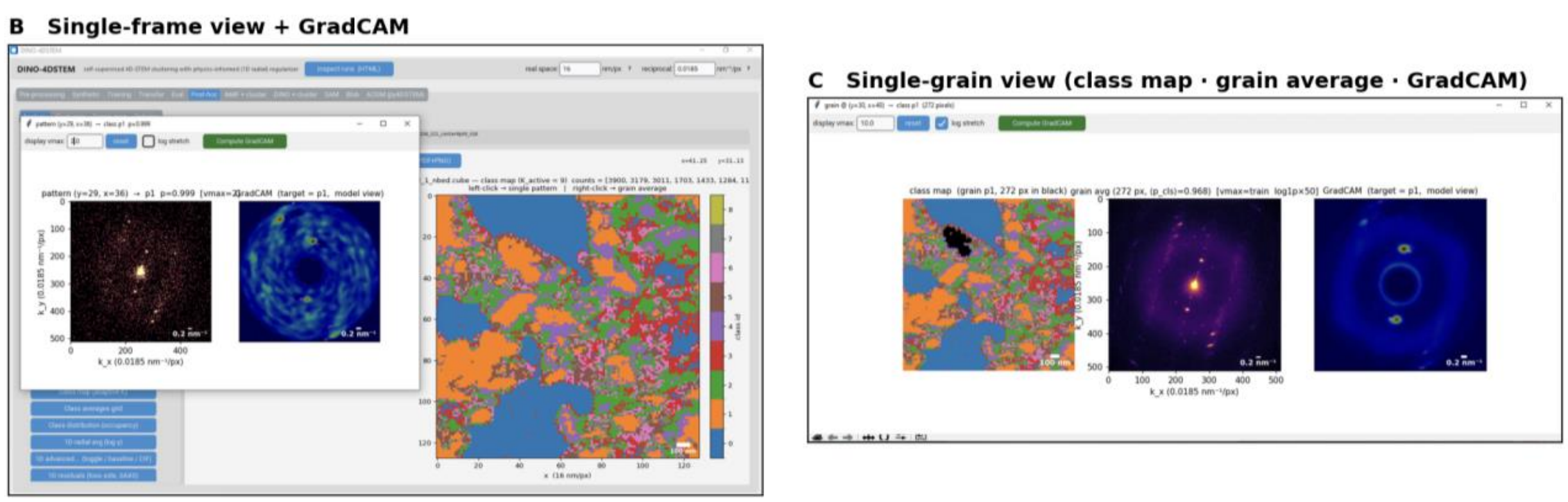

**Figure S14. Graphical interface and multi-scale inspection.** The DINO4DSTEM graphical interface. The user loads a dataset, runs the fixed-configuration model, and inspects the result at the level of a single frame, a grain, or a whole class, with gradient-based attribution (Grad-CAM) indicating which part of a pattern drove each assignment. A natural-language assistant drives the same workflow from plain-language requests.

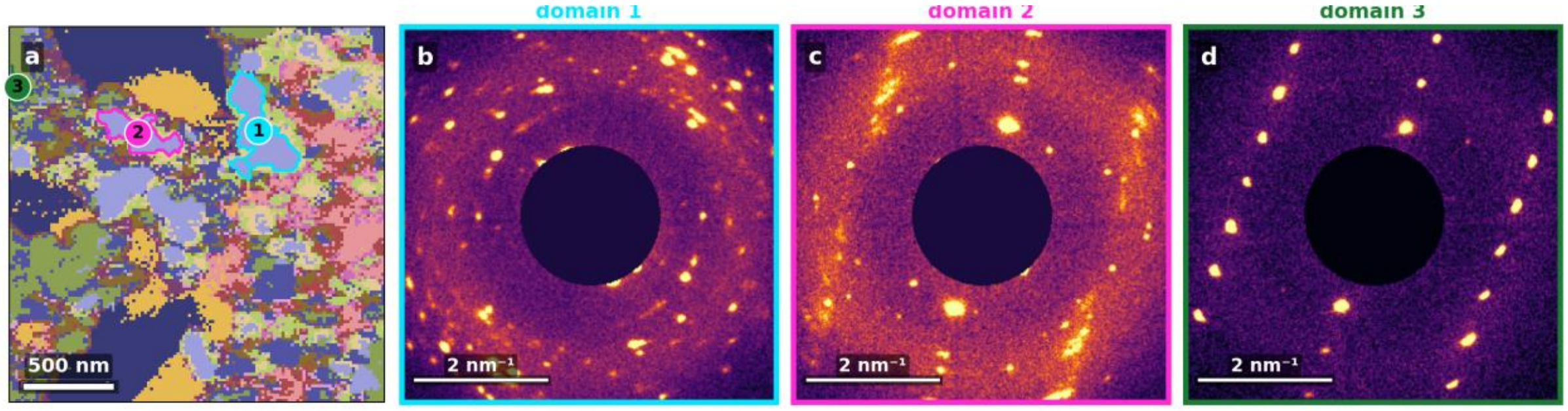

**Figure S15. Variation within a single class.** Three grains assigned to the same DINO4DSTEM class in the magnified interface field: the class map (a) with the three grains outlined, and their grain-average diffraction (b to d), ordered by increasing

crystalline order. Although all belong to one class, the Bragg reflections sharpen relative to the diffuse halo from b to d, showing that a single class still spans a small range of crystalline order and confirming that the classifier grades the data by degree of crystallization.

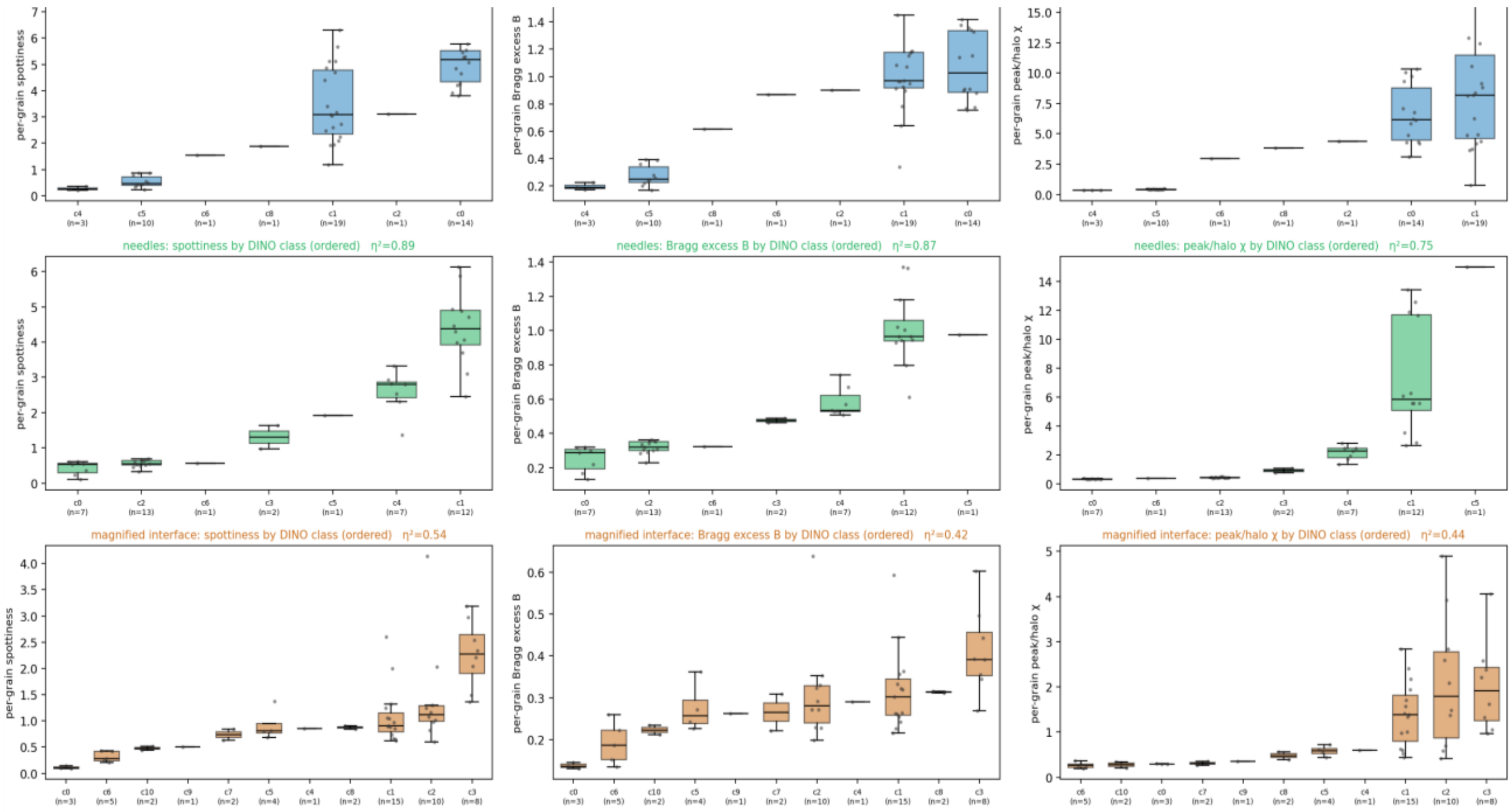

**Figure S16. Per-class distribution of the crystallization descriptors (indomethacin).** Each box is one DINO4DSTEM class; dots are its individual grains; classes are ordered by median. Rows are the three co-located fields (interface, needles, magnified interface; N = 49, 43, and 53 grains), columns the three rotation-invariant descriptors (azimuthal spottiness, two-dimensional Bragg excess B, radial peak-to-halo ratio χ). Measured per grain independently of the class label, the classes emerge as tight, separated, monotonic strata: the class label accounts for $\eta^2$ = 0.54 to 0.89 of the spottiness variance, 0.42 to 0.87 of B, and 0.44 to 0.75 of χ across the three fields. The class-median spottiness climbs continuously from about 0.2 (a near-uniform halo) to about 5 (discrete α Bragg spots). The three descriptors agree on the ordering and on the most crystalline classes but not always on their internal ranking, the fully crystalline classes differing in the character of their order rather than its amount.

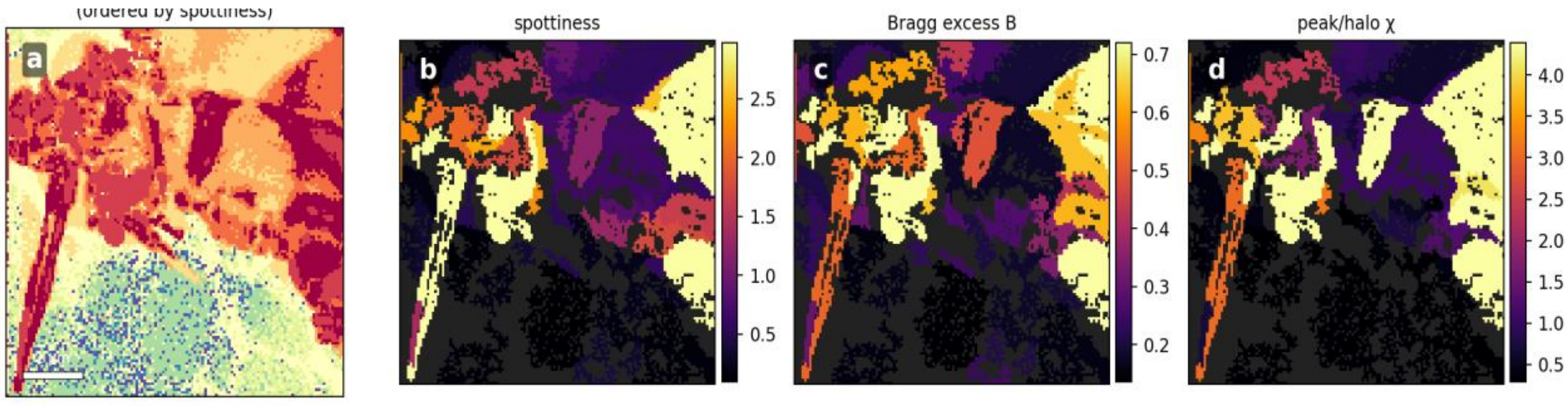

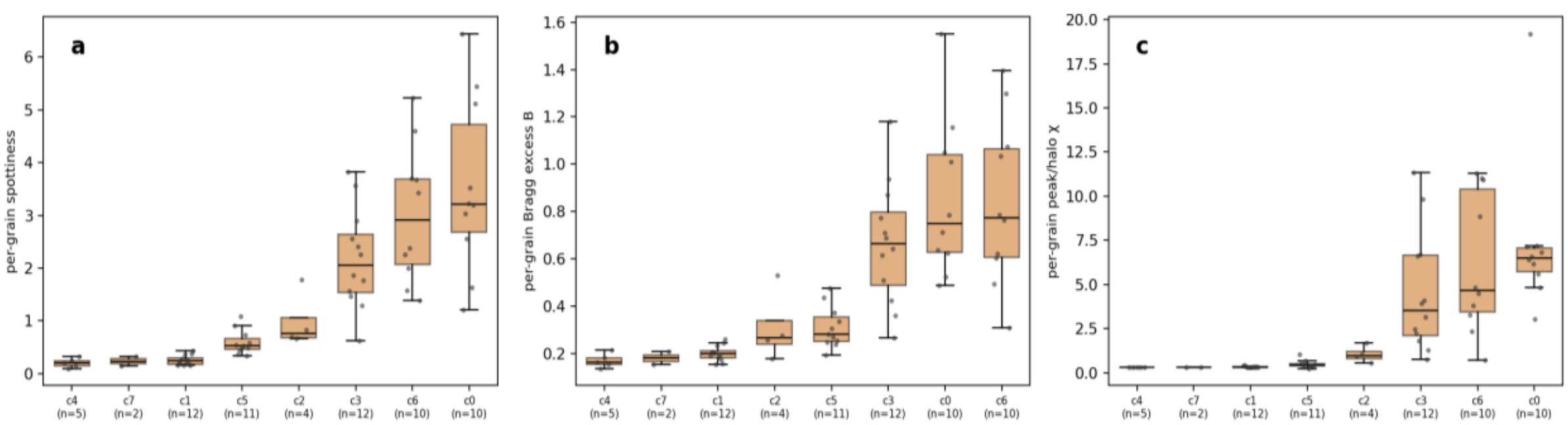

**Figure S17. Separate indomethacin needle field.** A separate needle field of the same annealed film, not used in the main analysis, processed identically. Top: the DINO4DSTEM class map and the three descriptor maps (as in Figure 6). Bottom: the per-class descriptor distributions (as in Figure S16). The same ordering ladder is reproduced (N = 66 grains, 8 classes, spottiness $\eta^2 = 0.68$, class-median spottiness 0.20 to 3.21), with the principal ring at d = 4.56 Å, confirming that the fixed-d azimuthal-sharpening signature is not specific to one field of view. As in the main fields, the descriptors agree on the ordering and on the top classes but not on their exact internal ranking, the fully crystalline classes differing in the character of their order rather than its amount.

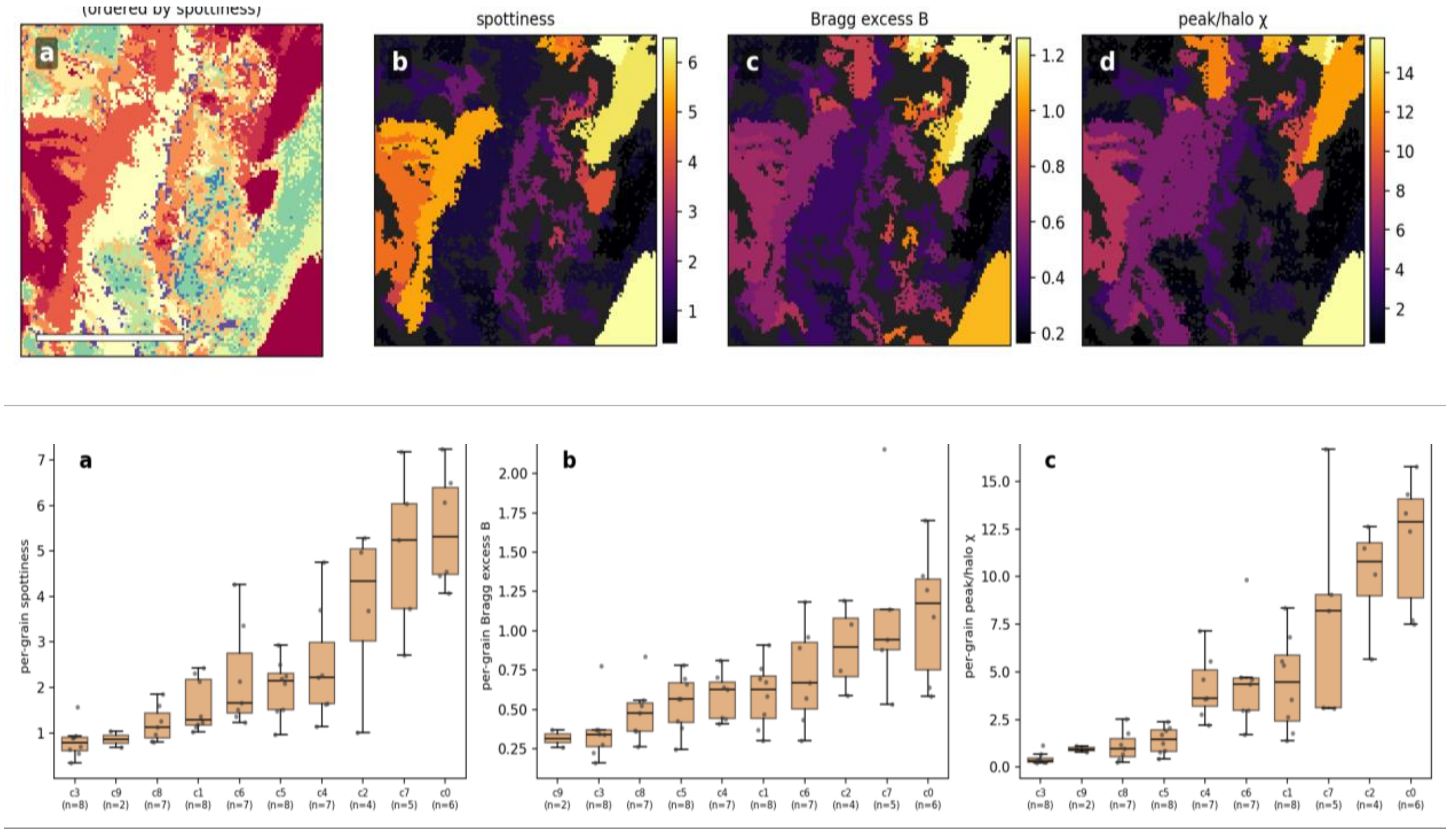

**Figure S18. Indomethacin thin-film region.** A thin-film indomethacin region processed identically (N = 62 grains, 10 classes, spottiness $\eta^2 = 0.71$, class-median spottiness 0.79 to 5.31; principal ring d = 4.67 Å). The same monotonic order axis is recovered, showing that the result holds across sample geometries. As in the main fields, the descriptors agree on the ordering and on the top classes but not on their exact internal ranking, the fully crystalline classes differing in the character of their order rather than its amount.

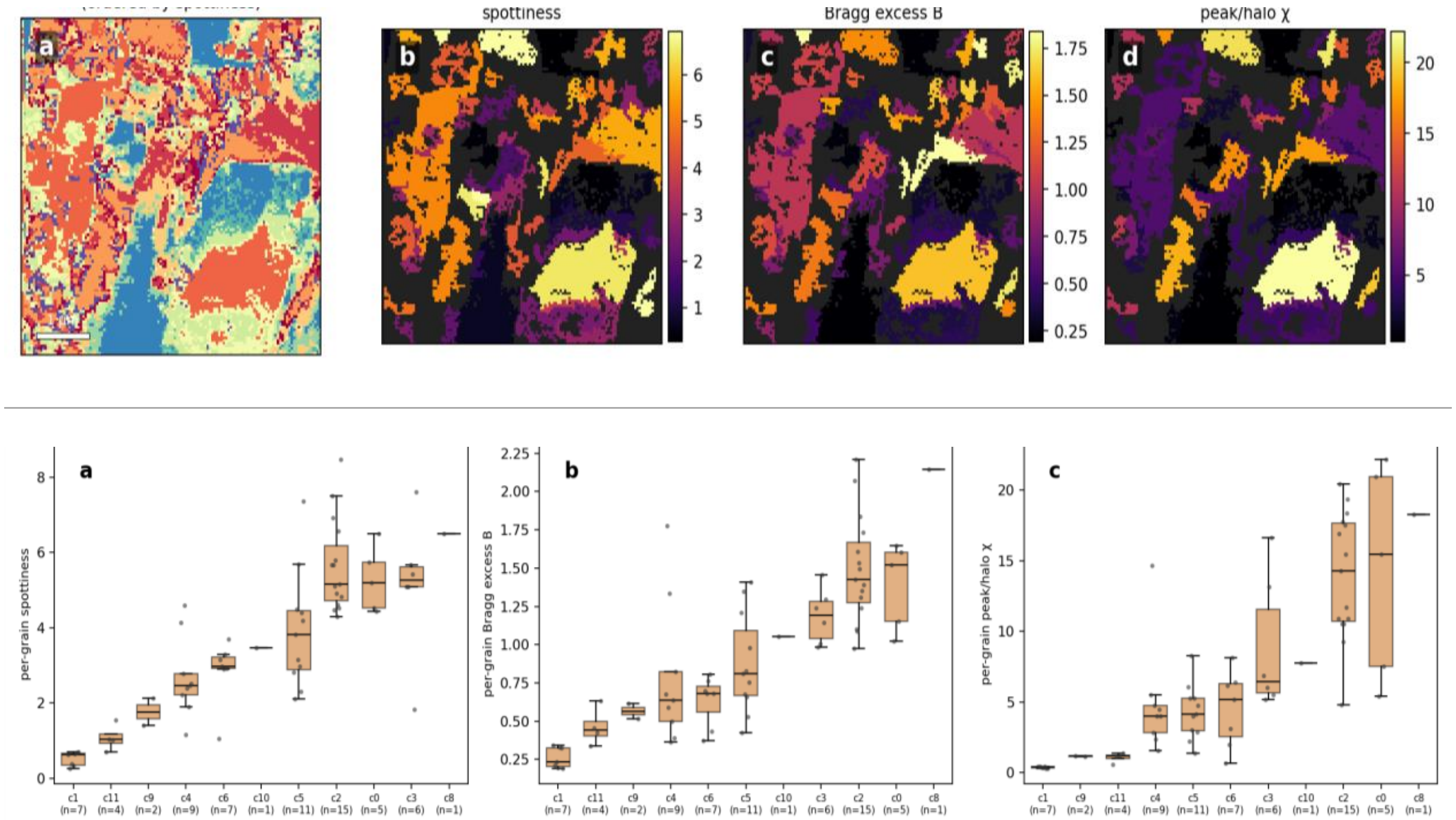

**Figure S19. Second independent indomethacin region.** A further indomethacin field (44 nm per pixel) processed identically (N = 68 grains, 11 classes, spottiness $\eta^2 = 0.72$, class-median spottiness 0.63 to 6.51; principal ring d = 4.72 Å, the closest of all fields to the nominal α {102}/{112} spacing). Top: the class map and three descriptor maps (as in Figure 6); bottom: the per-class distributions (as in Figure S16). The monotonic order axis and the fixed ring position are reproduced once more. As in the main fields, the descriptors agree on the ordering and on the top classes but not on their exact internal ranking, the fully crystalline classes differing in the character of their order rather than its amount.

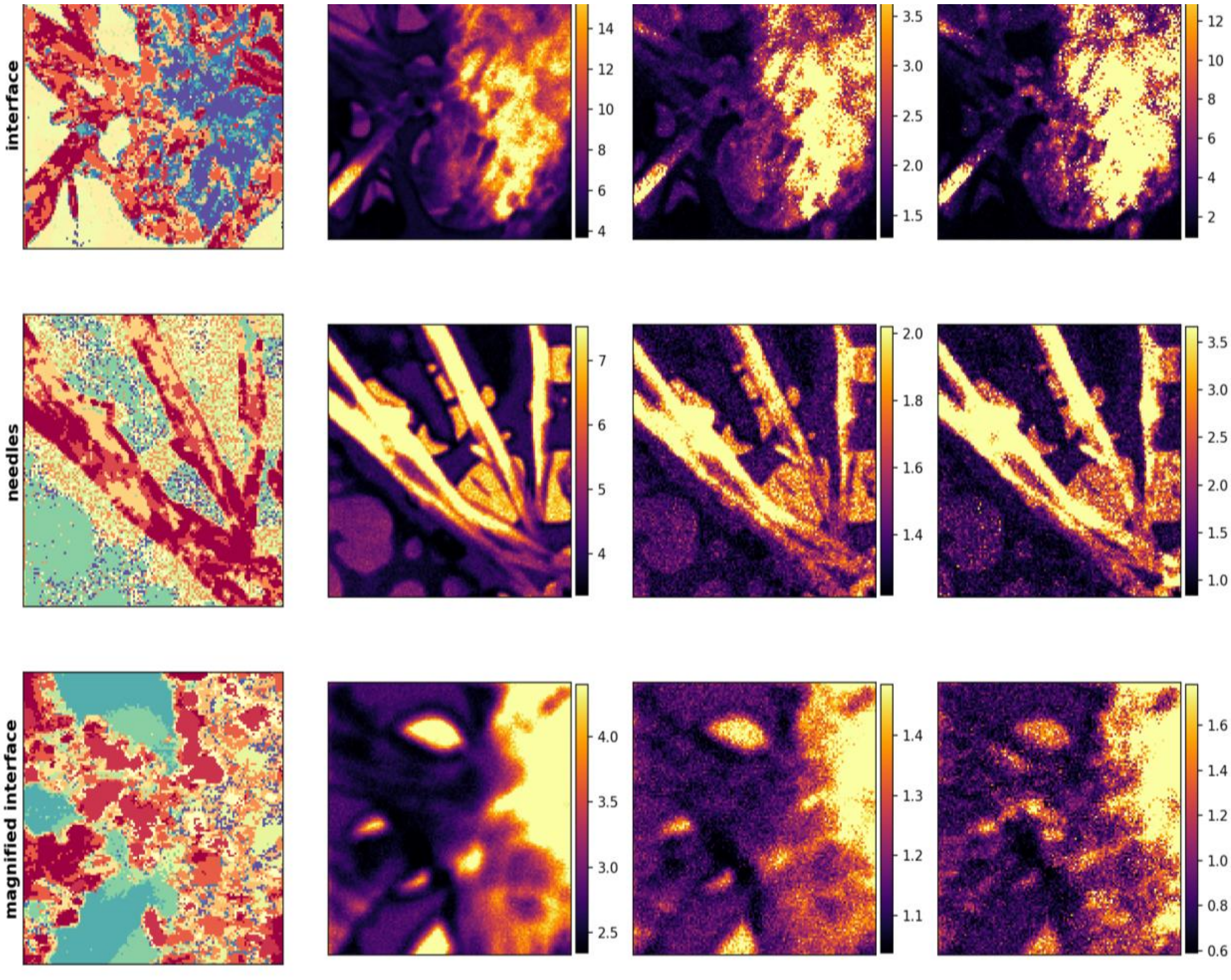
interface
needles
magnified interface
14
12
10
8
6
4
3.5
3.0
2.5
2.0
1.5
12
10
8
6
4
2
7
6
5
4
2.0
1.8
1.6
1.4
3.5
3.0
2.5
2.0
1.5
1.0
4.0
3.5
3.0
2.5
1.4
1.3
1.2
1.1
1.6
1.4
1.2
1.0
0.8
0.6

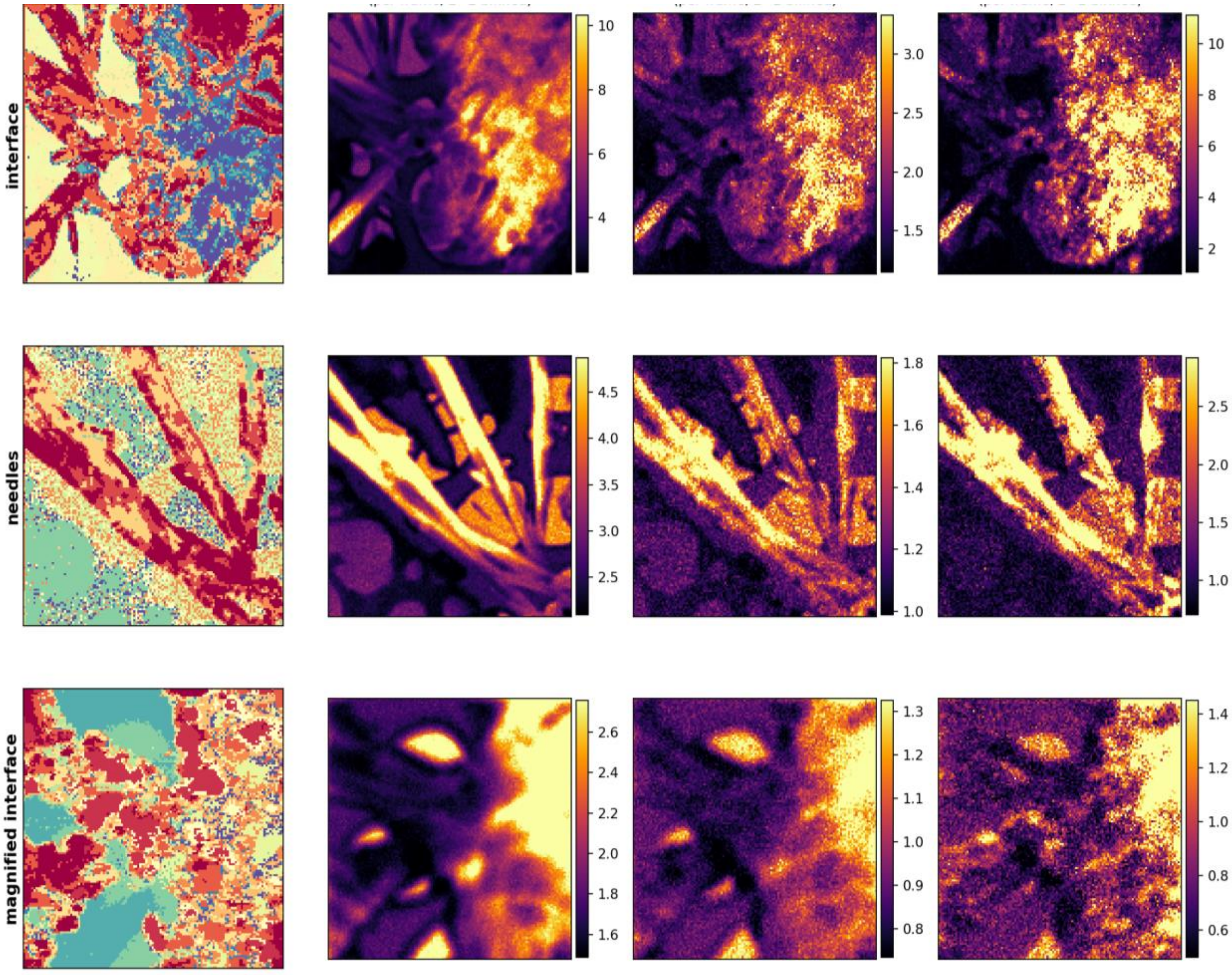

**Figure S20. The descriptors are meaningful only after clustering.** The same three descriptors measured on individual low-dose frames (top) and after 2×2 detector binning (bottom), rendered on the same maps as Figure 6. At the single-frame level shot noise dominates and the coherent spatial gradient of Figure 6 is absent; the apparent ordering is noisy and can locally invert. Binning does not recover the class-level behaviour, confirming that it is the clustering, not pixel-level averaging, that makes the order metrics measurable.